\documentclass[journal]{IEEEtran}
\usepackage{longtable}
\usepackage{cite}
\ifCLASSINFOpdf
  \usepackage[pdftex]{graphicx}
\else
\fi
\usepackage{array}
\usepackage{url}
\usepackage[hang]{subfigure}

\usepackage[normalem]{ulem} 

\usepackage{multirow}
\usepackage{booktabs}
\usepackage{makecell}

\usepackage{tikz}
\def\checkmark{\tikz\fill[scale=0.4](0,.35) -- (.25,0) -- (1,.7) -- (.25,.15) -- cycle;}

\newlength{\Oldarrayrulewidth}

\begin{document}
\title{Survey on Network Virtualization Hypervisors for
 Software Defined Networking}

\author{Andreas Blenk, 
        Arsany Basta, 
        Martin Reisslein,~\IEEEmembership{Fellow, IEEE,}
        and~Wolfgang Kellerer,~\IEEEmembership{Senior Member, IEEE}
\thanks{Please direct correspondence to M.~Reisslein.}
\thanks{A.~Blenk, A.~Basta, and W.~Kellerer are with the
Lehrstuhl f\"ur Kommunikationsnetze, Technische Universit\"at M\"unchen,
Munich, 80290, Germany,
(email: \{andreas.blenk, arsany.basta, wolfgang.kellerer\}@tum.de).}
\thanks{M.~Reisslein is with Elect., Comp., and Energy Eng.,
Arizona State Univ., Tempe, AZ 85287-5706, USA (email: reisslein@asu.edu),
phone 480-965-8593.}
}

\maketitle
\begin{abstract}
Software defined networking (SDN) has emerged as a promising
paradigm for making the control of communication networks flexible.
SDN separates the data packet forwarding plane, i.e., the data plane,
from the control plane and employs a central controller.
Network virtualization allows the flexible sharing of physical
networking resources by multiple users (tenants). Each tenant runs its own
applications over its virtual network, i.e., its slice of the actual
physical network.
The virtualization of SDN networks promises to allow
networks to leverage the combined benefits of SDN networking and
network virtualization and has therefore attracted significant research
attention in recent years.
A critical component for virtualizing SDN networks is an SDN hypervisor that
abstracts the underlying physical SDN network into multiple logically isolated
virtual SDN networks (vSDNs), each with its own controller.
We comprehensively survey hypervisors for SDN networks in this article.
We categorize the SDN hypervisors according to their architecture into
centralized and distributed hypervisors. We furthermore sub-classify
the hypervisors according to their execution platform into hypervisors
running exclusively on general-purpose compute platforms, or on a
combination of general-purpose compute platforms with general- or
special-purpose network elements.
We exhaustively compare the network attribute abstraction and isolation
features of the existing SDN hypervisors.
As part of the future research agenda, we
outline the development of a performance evaluation framework for
SDN hypervisors.
\end{abstract}

\begin{IEEEkeywords}
Centralized hypervisor, Distributed hypervisor, Multi-tenancy,
Network attribute abstraction, Network attribute isolation,
Network virtualization, Software defined networking.
\end{IEEEkeywords}

\IEEEpeerreviewmaketitle

\section{Introduction}\label{sec:int}

\subsection{Hypervisors: From Virtual Machines to Virtual Networks}
Hypervisors (also known as virtual machine monitors)
have initially been developed in the area of
virtual computing to monitor virtual
machines~\cite{gold1974sur,mal2014sur,ros2005vir}.
Multiple virtual machines can operate on a given computing platform.
For instance, with full virtualization, multiple virtual machines,
each with its own guest operating system, are running on a given (hardware,
physical) computing platform~\cite{kiv2007kvm,ros1999vmw,wal2002mem}.
Aside from monitoring the virtual machines, the hypervisor allocates
resources on the physical computing platform,
e.g., compute cycles on central processing units (CPUs),
to the individual virtual machines.
The hypervisor typically relies on an abstraction of the physical
computing platform, e.g., a standard instruction set,
for interfacing with the physical computing platform~\cite{smi2005arc}.
Virtual machines have become very important in computing as they
allow applications to flexibly run their operating systems
and programs without worrying about the specific details
and characteristics of the underlying computing platform, e.g.,
processor hardware properties.

Analogously, virtual networks have recently emerged to flexibly allow for
new network services without worrying about the specific details of
the underlying (physical) network, e.g., the specific underlying networking
hardware~\cite{ChBo09}.
Also, through virtualization, multiple virtual networks can flexibly
operate over the same physical network infrastructure.
In the context of virtual networking, a hypervisor monitors the
virtual networks and allocates networking resources, e.g.,
link capacity and buffer capacity in switching nodes, to the individual
virtual networks
(slices of the overall network)~\cite{Sherwood2009,Sherwood2010,kop2015net}.

Software defined networking (SDN) is a networking paradigm
that separates the control plane from the data (forwarding) plane,
centralizes the network control, and defines open, programmable
interfaces~\cite{Survey2014}.
The open, programmable interfaces allow for flexible interactions between
the networking applications and the underlying physical network
(i.e., the data plane) that is employed
to provide networking services to the applications.
In particular, the OpenFlow (OF) protocol~\cite{keo2008of}
provides a standardized
interface between the control plane and the underlying physical network
(data plane).
The OF protocol thus abstracts the underlying network, i.e., the
physical network that forwards the payload data.
The standardized data-to-control plane interface provided by the OF protocol
has made SDN a popular paradigm for network virtualization.

\subsection{Combining Network Virtualization and Software Defined
Networking}

Using the OF protocol, a hypervisor can establish
multiple \textit{virtual SDN networks (vSDNs)} based on a given
physical network.
Each vSDN corresponds to a ``slice'' of the overall network.
The virtualization of a given physical SDN network infrastructure
through a hypervisor allows multiple tenants (such as service providers
and other organizations) to share the SDN network infrastructure.
Each tenant can operate its own virtual SDN network, 
i.e., its own network operating system, independent
of the other tenants.

Virtual SDN networks are foreseen as enablers for future networking
technologies in fifth generation (5G) wireless
networks~\cite{metis5g2015,ngmn5g2015}.
Different services, e.g., voice and video, can run on isolated virtual slices
to improve the service quality and overall network 
performance~\cite{Anderson2005}.
Furthermore, virtual slices can
accelerate the development of new networking concepts by creating virtual
testbed environments.
For academic research, virtual SDN testbeds, e.g., GENI~\cite{ber2014gen} and
OFNLR~\cite{ofnlr},
offer the opportunity to easily create network slices on a short-term basis.
These network slices can be used to
test new networking concepts. In industry, new concepts can be
rolled out and tested in an isolated virtual SDN slice that can operate in
parallel with the operational (production) network.
This parallel operation can facilitate the roll out and at the same time
prevent interruptions of current network operations.

\subsection{Scope of this Survey}
This survey considers the topic area at the intersection of virtual
networking and SDN networking.
More specifically, we focus on hypervisors for virtual SDN networks, i.e.,
hypervisors for the creation and monitoring of and resource allocation to
virtual SDN networks. The surveyed hypervisors slice
a given physical SDN network into multiple vSDNs.
Thus, a hypervisor enables multiple tenants (organizational entities)
 to independently operate over a given
physical SDN network infrastructure, i.e., to run different network 
operating systems in parallel.

\subsection{Contributions and Structure of this Article}
This article provides a comprehensive survey of hypervisors for
virtual SDN networks. We first provide brief tutorial background and
review existing surveys on the related topic areas of network
virtualization and software defined networking (SDN) in
Section~\ref{relwork:sec}. The main acronyms used in this article
are summarized in Table~\ref{acro:tab}. 
In Section~\ref{virsdnnet:sec}, we introduce the virtualization of SDN
networks through hypervisors and describe the two main hypervisor
functions, namely the virtualization (abstraction) of network
attributes and the isolation of network attributes. In
Section~\ref{class:sec}, we introduce a classification of
hypervisors for SDN networks according to their architecture as
centralized or distributed hypervisors. We further sub-classify the
hypervisors according to their execute platform into hypervisors
implemented through software programs executing on general-purpose
compute platforms or a combination of general-purpose compute
platforms with general- or special-purpose network elements (NEs).
We then survey the existing hypervisors following the introduced
classification: centralized hypervisors are surveyed in
Section~\ref{central:sec}, while distributed hypervisors are
surveyed in Section~\ref{distr:sec}.
We compare the surveyed hypervisors in Section~\ref{sec:summary},
whereby we contrast in particular the abstraction (virtualization)
of network attributes and the isolation of vSDNs.
In Section~\ref{perf:sec}, we survey the existing performance
evaluation tools, benchmarks, and evaluation scenarios for
SDN networks and introduce a framework for
the comprehensive performance evaluation of SDN hypervisors.
Specifically, we initiate the establishment of a sub-area of SDN
hypervisor research by defining SDN hypervisor performance metrics and
specifying performance evaluation guidelines for SDN hypervisors.
In Section~\ref{fut:sec}, we outline an agenda for future
research on SDN hypervisors.
We conclude this survey article in Section~\ref{concl:sec}.

\section{Background and Related Surveys}
\label{relwork:sec}
In this section, we give tutorial background on the topic areas of
network virtualization and software defined networking (SDN).
We also give an overview of existing survey articles in these topic areas.

\subsection{Network Virtualization}\label{relwork:secA}
Inspired by the success of virtualization in the area of
computing~\cite{gold1974sur,li2010sur,sah2010vir, Douglis2013a,smi2005arc,zha2014vmt},
virtualization has become an important research topic in communication networks.
Initially, network virtualization was conceived to ``slice''
a given physical network infrastructure into multiple virtual networks,
also referred to as
``slices''~\cite{ahm2015tow,ber2014gen,etsi_nfv002,han2003xorp,vil2001vir}.
Network virtualization first abstracts
the underlying physical network and then creates separate
virtual networks (slices) through specific abstraction and isolation
functional blocks that are reviewed in detail in Section~\ref{virsdnnet:sec}
(in the context of SDN).
Cloud computing platforms and their virtual service functions have been
surveyed in~\cite{bist2013comp,endo2010sur,soa2015tow},
while related security issues have been surveyed in~\cite{sub2011sur}.

\begin{center}
\begin{table}[t]
\centering
\caption{Summary of main acronyms}
\label{acro:tab}
\begin{tabular}{|l|l|} \hline
A-CPI & Application-Controller Plane Interface~\cite{SDNarch10,SDNarch11} \\
API  & Application Programmers Interface \\
D-CPI  & Data-Controller Plane Interface~\cite{SDNarch10,SDNarch11}\\
IETF  & Internet Engineering Task Force \\
MPLS & Multiple Protocol Label Switching~\cite{xia2000tra} \\
NE   & Network Element \\
OF  & OpenFlow~\cite{Hu2014,LaKR14,OF15} \\
SDN & Software Defined Networking, or \\
    & \ \ \ Software Defined Network, \\
    & \ \ \ depending on context \\
TCAM & Ternary Content Addressable Memory~\cite{pan2002red}\\
VLAN & Virtual Local Area Network \\
vSDN & virtual Software Defined Network\\
WDM  & Wavelength Division Multiplexing \\
\hline
\end{tabular}
\end{table}
\end{center}

In the networking domain, there are several related techniques that
can create network ``slices''. For instance,
wavelength division multiplexing (WDM)~\cite{ish1984rev} creates
slices at the physical (photonic) layer, while
virtual local area networks (VLANs)~\cite{yu2011sur} create slices at
the link layer.
Multiple protocol label switching (MPLS)~\cite{xia2000tra} creates slices of
forwarding tables in switches.
In contrast, network virtualization seeks to create slices
of the entire network, i.e., to form virtual networks (slices) across
all network protocol layers.
A given virtual network (slice) should have its own resources, including
its own slice-specific view of the network topology, its own
slices of link bandwidths, and its own slices of switch CPU resources and
switch forwarding tables.

A given virtual network (slice) provides a setting for examining
novel networking paradigms, independent of the constraints imposed
by the presently dominant Internet structures and protocols~\cite{Anderson2005}.
Operating multiple virtual networks over a given network
infrastructure with judicious resource allocation
may improve the utilization of the networking
hardware~\cite{bel2012res,leo2003vir}.
Also, network virtualization allows multiple network service providers
to flexibly offer new and innovative services over an existing
underlying physical network
infrastructure~\cite{Chowdhury2008,ChBo09,hwa2015net,Khan2012}.

The efficient and reliable operation of virtual networks typically demands
specific amounts of physical networking resources. In order to
  efficiently utilize the resources of the virtualized
  networking infrastructure,
  sophisticated resource allocation algorithms are needed to assign
  the physical resources to the virtual networks. Specifically,
   the virtual nodes
  and the virtual paths that interconnect the virtual nodes
  have to be placed on the physical infrastructure. This assignment
  problem of virtual resources to physical resources is known as
  the Virtual Network Embedding (VNE)
problem~\cite{Blenk2013,cho2012vin,fis2013vir}. The VNE problem is
NP-hard and is still intensively studied. Metrics to quantify and
compare embedding algorithms include acceptance rate, the revenue
per virtual network, and the cost per virtual network. The
acceptance rate is defined as the ratio of the number of accepted
virtual networks to the total number of virtual network requests. If
the resources are insufficient for hosting a virtual network, then
the virtual network request has to be rejected.  Accepting and
rejecting virtual network requests has to be implemented by
admission control mechanisms. The revenue defines the gain per
virtual network, while the cost defines the physical resources that
are expended to accept a virtual network. The revenue-to-cost ratio
provides a metric relating both revenue and cost; a high
revenue-to-cost ratio indicates an efficient embedding. Operators
may select for which metrics the embedding of virtual networks
should be optimized. The existing VNE algorithms, which range from
exact formulations, such as mixed integer linear programs, to
heuristic approaches based, for instance, on stochastic sampling,
have been surveyed in~\cite{fis2013vir}. Further,~\cite{fis2013vir}
outlines metrics and use cases for VNE algorithms. In
Section~\ref{perf:sec}, we briefly relate the assignment (embedding)
of vSDNs to the generic VNE problem and outline the use of the
general VNE performance metrics in the SDN context.

Several overview articles and surveys have addressed the
general principles, benefits, and mechanisms of network
virtualization~\cite{bari2013data,Car2009NVV,Fer2011,gho2014tow,han2015net,mij2015net,pen2015gue,ryg2013net,she2015vcon}.
An instrumentation and analytics framework for virtualized networks
has been outlined in~\cite{vei2015ins} while convergence mechanisms
for networking and cloud computing have been surveyed
in~\cite{duan2012sur,soa2015tow}.
Virtualization in the context of
wireless and mobile networks is an emerging area that
has been considered
in~\cite{haw2014nfv, jar2015sdn, kha2015wir, li 2012tow, Li, lia2015wirCOMST, lia2015wir, Nguyen2015, wan2013wir, wen2013wir}.

\begin{figure*}[t!]
\setlength{\unitlength}{1mm}
    \centering
    \subfigure[Conventional SDN Network: An SDN controller directly
interacts with the physical SDN network to provide services to the
applications.]{
\begin{picture}(56,85)
\put(0,0){\makebox(0,0)[lb]{\framebox(50,45)}}
\put(5,5){\circle{2}} \put(5,20){\circle{2}} \put(5,35){\circle{2}}
\put(25,5){\circle{2}} \put(25,35){\circle{2}} \put(45,20){\circle{2}}
\put(6,5){\line(1,0){18}} \put(6,35){\line(1,0){18}}
\put(5,6){\line(0,1){13}} \put(5,21){\line(0,1){13}}
\put(25,6){\line(0,1){28}} \put(25.7,5.7){\line(4,3){18.6}}
\put(25.7,34.3){\line(4,-3){18.6}}
\put(4,37.5){\makebox(0,0)[lb]{\shortstack[c]{Physical SDN Network}}}
\put(10,60){\makebox(0,0)[lb]{\framebox(30,7)}}
\put(24,52){\makebox(0,0)[br]{\shortstack[c]{D-CPI}}}
\put(25,62){\makebox(0,0)[b]{\shortstack[c]{SDN Controller}}}
\put(3,73){\makebox(0,0)[lb]{\framebox(18,7)}}
\put(12,75){\makebox(0,0)[b]{\shortstack[c]{$\mbox{App}_1$}}}
\put(29,73){\makebox(0,0)[lb]{\framebox(18,7)}}
\put(38,75){\makebox(0,0)[b]{\shortstack[c]{$\mbox{App}_2$}}}
\put(11,68.5){\makebox(0,0)[br]{\shortstack[c]{A-CPI}}}
\put(37,68.5){\makebox(0,0)[br]{\shortstack[c]{A-CPI}}}
\put(12,70){\vector(0,-1){3}}  \put(12,70){\vector(0,1){3}}
\put(38,70){\vector(0,-1){3}}  \put(38,70){\vector(0,1){3}}
\put(25,52.5){\vector(0,-1){7.5}}  \put(25,52.5){\vector(0,1){7.5}}
\end{picture} }
    \subfigure[Virtualizing the SDN Network (Perspective of the
hypervisor): The inserted hypervisor
directly interacts with the physical SDN Networks as well as
two virtual SDN controllers.]{
\begin{picture}(56,85)
\put(0,0){\makebox(0,0)[lb]{\framebox(50,45)}}
\put(5,5){\circle{2}} \put(5,20){\circle{2}} \put(5,35){\circle{2}}
\put(25,5){\circle{2}} \put(25,35){\circle{2}} \put(45,20){\circle{2}}
\put(6,5){\line(1,0){18}} \put(6,35){\line(1,0){18}}
\put(5,6){\line(0,1){13}} \put(5,21){\line(0,1){13}}
\put(25,6){\line(0,1){28}} \put(25.7,5.7){\line(4,3){18.6}}
\put(25.7,34.3){\line(4,-3){18.6}}
\put(4,37.5){\makebox(0,0)[lb]{\shortstack[c]{Physical SDN Network}}}
\put(7.5,49){\makebox(0,0)[lb]{\framebox(35,7)}}
\put(25,51){\makebox(0,0)[b]{\shortstack[c]{Hypervisor}}}
\put(25,47){\vector(0,-1){2}}  \put(25,47){\vector(0,1){2}}
\put(24,46){\makebox(0,0)[br]{\shortstack[c]{{\footnotesize{D-CPI}}}}}
\put(0,60){\makebox(0,0)[lb]{\framebox(23,7)}}
\put(11.5,62){\makebox(0,0)[b]{\shortstack[c]{vSDN 1 Contr.}}}
\put(11.5,58){\vector(0,-1){2}}  \put(11.5,58){\vector(0,1){2}}
\put(10.5,57){\makebox(0,0)[br]{\shortstack[c]{{\footnotesize{D-CPI}}}}}
\put(27,60){\makebox(0,0)[lb]{\framebox(23,7)}}
\put(38.5,62){\makebox(0,0)[b]{\shortstack[c]{vSDN 2 Contr.}}}
\put(38.5,58){\vector(0,-1){2}}  \put(38.5,58){\vector(0,1){2}}
\put(37.5,57){\makebox(0,0)[br]{\shortstack[c]{{\footnotesize{D-CPI}}}}}
\put(0,73){\makebox(0,0)[lb]{\framebox(10,7)}}
\put(5,75){\makebox(0,0)[b]{\shortstack[c]{$\mbox{App}_{11}$}}}
\put(5,70){\vector(0,-1){3}}  \put(5,70){\vector(0,1){3}}
\put(4,69){\makebox(0,0)[br]{\shortstack[c]{{\small{A-CPI}}}}}
\put(13,73){\makebox(0,0)[lb]{\framebox(10,7)}}
\put(18,75){\makebox(0,0)[b]{\shortstack[c]{$\mbox{App}_{12}$}}}
\put(18,70){\vector(0,-1){3}}  \put(18,70){\vector(0,1){3}}
\put(17,69){\makebox(0,0)[br]{\shortstack[c]{{\small{A-CPI}}}}}
\put(27,73){\makebox(0,0)[lb]{\framebox(10,7)}}
\put(32,75){\makebox(0,0)[b]{\shortstack[c]{$\mbox{App}_{21}$}}}
\put(32,70){\vector(0,-1){3}}  \put(32,70){\vector(0,1){3}}
\put(31,69){\makebox(0,0)[br]{\shortstack[c]{{\small{A-CPI}}}}}
\put(40,73){\makebox(0,0)[lb]{\framebox(10,7)}}
\put(45,75){\makebox(0,0)[b]{\shortstack[c]{$\mbox{App}_{22}$}}}
\put(45,70){\vector(0,-1){3}}  \put(45,70){\vector(0,1){3}}
\put(44,69){\makebox(0,0)[br]{\shortstack[c]{{\small{A-CPI}}}}}
\end{picture} }
    \subfigure[Virtual SDN Networks (Perspective of the
virtual SDN controllers): Each virtual SDN controller
has the perception of transparently interacting
with its virtual SDN network.]{
\begin{picture}(56,85)
\put(0,0){\makebox(0,0)[lb]{\framebox(23,45)}}
\put(2.5,20){\circle{2}}
\put(11.5,5){\circle{2}} \put(11.5,35){\circle{2}} \put(20.5,20){\circle{2}}
\put(11.5,6){\line(0,1){28}}
\put(11.7,5.7){\line(3,5){8.3}}
\put(11.7,34.3){\line(3,-5){8.3}}
\multiput(2.5,5)(2,0){4}{\line(1,0){1.25}}
\multiput(2.5,5)(0,2){7}{\line(0,1){1.25}}
\multiput(2.5,35)(2,0){4}{\line(1,0){1.25}}
\multiput(2.5,35)(0,-2){7}{\line(0,-1){1.25}}
\put(2,37.5){\makebox(0,0)[lb]{\shortstack[c]{virtual SDN \\ Network 1}}}
\put(27,0){\makebox(0,0)[lb]{\framebox(23,45)}}
\put(29.5,5){\circle{2}} \put(29.5,20){\circle{2}} \put(29.5,35){\circle{2}}
\put(38.5,5){\circle{2}} \put(38.5,35){\circle{2}} \put(47.5,20){\circle{2}}
\put(30.5,35){\line(1,0){7}}
\put(29.5,6){\line(0,1){13}} \put(29.5,21){\line(0,1){13}}
\put(38.5,6){\line(0,1){28}}  \put(38.7,5.7){\line(3,5){8.3}}
\put(29,37.5){\makebox(0,0)[lb]{\shortstack[c]{virtual SDN \\ Network 2}}}
\put(7.5,49){\makebox(0,0)[lb]{\dashbox{2}(35,7)}}
\put(25,51){\makebox(0,0)[b]{\shortstack[c]{Hypervisor}}}
\put(0,60){\makebox(0,0)[lb]{\framebox(23,7)}}
\put(11.5,62){\makebox(0,0)[b]{\shortstack[c]{vSDN 1 Contr.}}}
\put(11.5,58){\vector(0,-1){13}}  \put(11.5,58){\vector(0,1){2}}
\put(27,60){\makebox(0,0)[lb]{\framebox(23,7)}}
\put(38.5,62){\makebox(0,0)[b]{\shortstack[c]{vSDN 2 Contr.}}}
\put(38.5,58){\vector(0,-1){13}}  \put(38.5,58){\vector(0,1){2}}
\put(0,73){\makebox(0,0)[lb]{\framebox(10,7)}}
\put(5,75){\makebox(0,0)[b]{\shortstack[c]{$\mbox{App}_{11}$}}}
\put(5,70){\vector(0,-1){3}}  \put(5,70){\vector(0,1){3}}
\put(13,73){\makebox(0,0)[lb]{\framebox(10,7)}}
\put(18,75){\makebox(0,0)[b]{\shortstack[c]{$\mbox{App}_{12}$}}}
\put(18,70){\vector(0,-1){3}}  \put(18,70){\vector(0,1){3}}
\put(27,73){\makebox(0,0)[lb]{\framebox(10,7)}}
\put(32,75){\makebox(0,0)[b]{\shortstack[c]{$\mbox{App}_{21}$}}}
\put(32,70){\vector(0,-1){3}}  \put(32,70){\vector(0,1){3}}
\put(40,73){\makebox(0,0)[lb]{\framebox(10,7)}}
\put(45,75){\makebox(0,0)[b]{\shortstack[c]{$\mbox{App}_{22}$}}}
\put(45,70){\vector(0,-1){3}}  \put(45,70){\vector(0,1){3}}
\end{picture} }
    \caption{Conceptual illustration of a conventional
physical SDN network and its virtualization:
In the conventional (non-virtualized) SDN network illustrated in part (a),
network applications ($\mbox{App}_1$ and $\mbox{App}_2$)
interact through the application-controller plane
interface (A-CPI) with the SDN controller, which in turn interacts through
the data-controller plane interface (D-CPI) with the physical SDN network.
The virtualization of the SDN network, so that the network can be shared by
two tenants is conceptually illustrated in parts (b) and (c).
A hypervisor is inserted between the physical SDN network and the SDN
controller (control plane). The hypervisor directly interacts
with the physical SDN network, as illustrated in part (b).
The hypervisor gives each virtual SDN (vSDN) controller the perception
that the vSDN controller directly (transparently) interacts with the
corresponding virtual SDN network, as illustrated in part (c).
The virtual SDN networks
are isolated from each other and may have different levels of abstraction and
different topology, as illustrated in part~(c),
although they are both based on the one physical SDN
network illustrated in part (b).}
\label{concepts:fig}
\end{figure*}
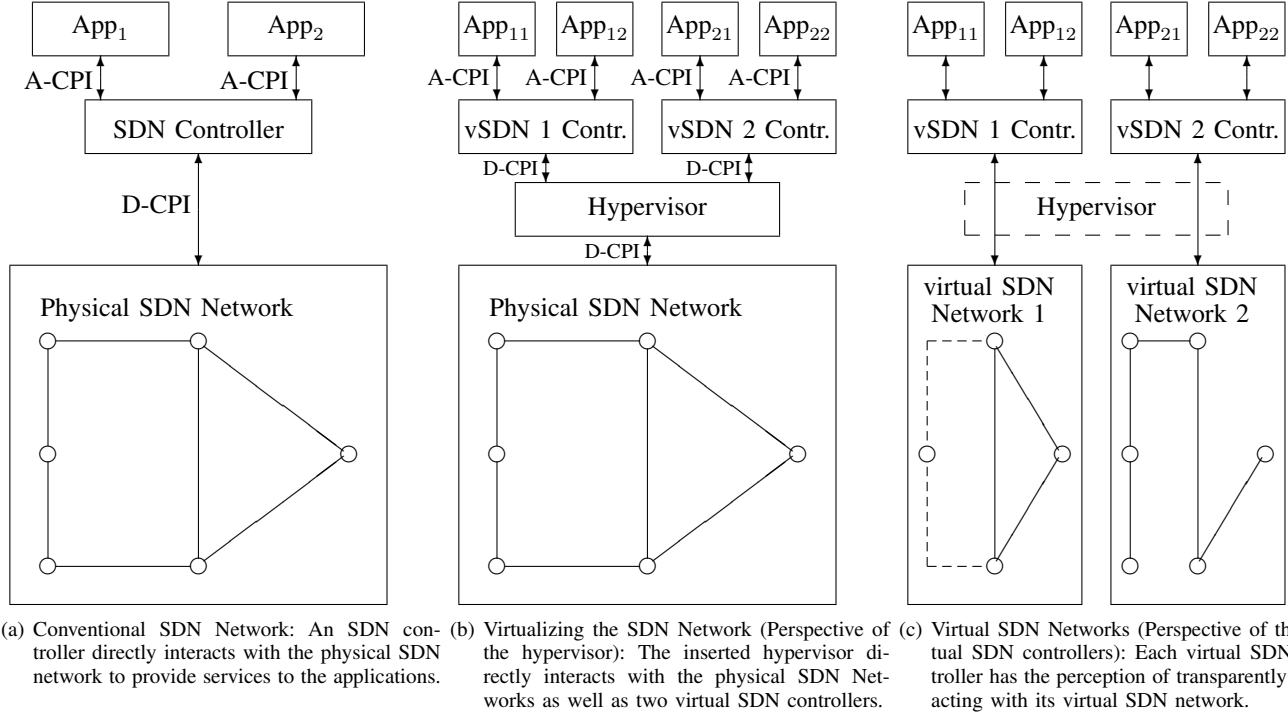
\normalsize
\subsection{Software Defined Networking}
\label{bg_sdn:sec}
Software Defined Networking (SDN) decouples the control plane, which
controls the operation of the network switches, e.g., by
setting the routing tables,
from the data (forwarding) plane, which carries out the actual
forwarding of the payload data through the physical network of
switches and links.

SDN breaks with the traditionally distributed control of the
Internet by centralizing the control of an  entire
physical SDN network in a single logical SDN
controller~\cite{Survey2014}, as illustrated in Fig.~\ref{concepts:fig}(a).
The first SDN controller based on the OpenFlow protocol
was NOX~\cite{gud2008nox,ToGGC12} and has been followed
by a myriad of controller designs,
e.g.,~\cite{eri2013bea,fos2011fre,guh2013mac,mon2012com,ryu,voe2012sca,Berde2014,opendaylight},
which are written in a variety of programming languages.
Efforts to distribute the SDN control plane decision making
to local controllers that are ``passively'' synchronized to maintain
central logical control are reported in~\cite{too2010hyp}.
Similar efforts to control large-scale SDN network have been
examined in~\cite{Bozakov2013,cho2014suv,dix2013tow,kwa2015rao},
while hybrid SDN networks
combining centralized SDN with traditional distributed protocols
are outlined in~\cite{vis2014opp}.

SDN defines a so-called data-controller plane interface (D-CPI)
between the physical data (forwarding)
plane and the SDN control plane~\cite{SDNarch10,SDNarch11}.
This interface has also been referred to as the
south-bound application programmers interface (API) or the
OF control channel or the control plane channel
in some SDN literature, e.g.,~\cite{JaZH14}.
The D-CPI relies on a standardized
instruction set that abstracts the physical data forwarding hardware.
The OF protocol~\cite{keo2008of} (which employs some aspects from
Orphal~\cite{mog2008orp}) is a widely employed SDN
instruction set, an alternative developed by the IETF is the
forwarding and control element separation (ForCES) 
protocol~\cite{dor2010for,hal2015net}.

The SDN controller interfaces through the application-controller
plane interface (A-CPI)~\cite{SDNarch10,SDNarch11},
which has also been referred to as north-bound API~\cite{JaZH14}, with
the network applications.
The network applications, illustrated by $\mbox{App}_1$ and $\mbox{App}_2$
in Fig.~\ref{concepts:fig}(a), are sometimes combined in a so-called
application control plane.
Network applications can be developed upon functions that are implemented
through SDN controllers.
Network application can implement network decisions or traffic steering.
Example network applications are firewall, router, network monitor, and
load balancer.
The data plane of the SDN that is controlled through an SDN controller can
then support the classical end-user network applications, 
such as the web
(HTTP) and e-mail (SMTP)~\cite{kur2012com,for2012com}.

The SDN controller interfaces through the
intermediate-controller plane interface (I-CPI) with the controllers
in other network domains.
The I-CPI includes the formerly defined east-bound API, that interfaces with
the control planes of network domains that are not
running SDN, e.g., with the MPLS control plane in a
non-SDN network domain~\cite{JaZH14}.
The I-CPI also includes the formerly defined west-bound API, that interfaces
with the SDN controllers in different network
domains~\cite{FuBG14,jai2013b4,phe2014dis,YeG14}.

SDN follows a match and action paradigm. The instruction set
pushed by the controller to the SDN data plane includes a
match specification that defines a packet flow.
This match can be specified based on values in the packet, e.g., values
of packet header fields. The OF protocol specification defines a set of
fields that an OF controller can use and an OF NE can match on.
This set is generally referred to as the ``flowspace''.
The OF match set, i.e., flowspace, is updated and extended with newer
versions of the OF protocol specification.
For example, OF version 1.1 extended OF version 1.0 with a match
capability on MPLS labels.
The controller instruction
set also includes the action to be taken by the data plane
once a packet has been matched, e.g., forward to a certain port or queue,
drop, or modify the packet. The action set is also continuously updated
and extended with newer OF specifications.

Several survey articles have covered the general principles of
SDN~\cite{dha2013sdn,far2015soft,FeRZ14,Hakiri2014,Hu2014,Jammal2014,jar2014sur,LaKR14,Survey2014,NuMN14,Xie2015,XiWF14}.
The implementation of software components for SDN has been surveyed
in~\cite{al2014sur,keo2008of,vau2011of}, while
SDN network management has been surveyed in~\cite{hel2013lev,kha2015jons,wic2015sof}.
SDN issues specific to cloud computing, optical, as well as mobile and
satellite networks have been covered
in~\cite{azo2013sdn,ban2013mer,bha2014soft,ber2015sof,bou2011of,cha2014joi,czi2014sdn,Networking2013,kna2014slo,li2014sof,Liu2014,mam2015ser,may2015exp,man2014het,MiCK13,Racherla,rag2012dyn,yan2014sof}.
SDN security has been surveyed 
in~\cite{ahm2015sec,als2015sec,sco2013sdn,sco2015sur,yan2015sof} 
while SDN scalability
and resilience has been surveyed in~\cite{kim2015sdn,Longo2015,van2014sca}.

\section{Virtualizing SDN Networks} \label{virsdnnet:sec}
In this section, we explain how to virtualize SDN networks through a
hypervisor.
We highlight the main functions that are implemented by a hypervisor
to create virtual SDN networks.

\subsection{Managing a Physical SDN Network with a Hypervisor}
The centralized SDN controller in conjunction with the
outlined interfaces (APIs) result in a ``programmable'' network.
That is, with SDN, a network is no longer a conglomerate of individual
devices that require individualized configuration
Instead, the SDN network can be viewed and operated as a single
programmable entity.
This programmability feature of SDN can be employed for
implementing virtual SDN networks (vSDNs)~\cite{Jain2013}.

More specifically, an SDN network can be virtualized by inserting
a hypervisor between the physical SDN network and the SDN control plane,
as illustrated in
Fig.~\ref{concepts:fig}(b)~\cite{Sherwood2009,Sherwood2010,kop2015net}.
The hypervisor views and interacts with the entire physical SDN
network through the D-CPI interface.
The hypervisor also interacts through multiple D-CPI interfaces with
multiple virtual SDN controllers.
We consider this interaction of a hypervisor via multiple D-CPIs with
multiple vSDN controllers, which can be conventional legacy SDN
controllers, as the defining feature of a hypervisor.
We briefly note that some controllers for non-virtualized SDN networks,
such as OpenDaylight~\cite{opendaylight}, could also provide
network virtualization.
However, these controllers would allow the control of virtual network
slices only via the A-CPI.
Thus, legacy SDN controllers  could not communicate transparently with
their virtual network slices.
We do therefore not consider OpenDaylight or similar controllers
as hypervisors.

The hypervisor abstracts (virtualizes) the physical SDN network
and creates isolated virtual SDN networks that are controlled
by the respective virtual SDN controllers.
In the example illustrated in Fig.~\ref{concepts:fig},
the hypervisor slices (virtualizes)
the physical SDN network in Fig.~\ref{concepts:fig}(b)
to create the two vSDNs illustrated in Fig.~\ref{concepts:fig}(c).
Effectively, the hypervisor abstracts its view of the
underlying physical SDN network in Fig.~\ref{concepts:fig}(b), into
the two distinct views of the two vSDN controllers in
Fig.~\ref{concepts:fig}(c).
This abstraction is sometimes referred to as an $n$-to-1 mapping 
or a many-to-one mapping in that the
abstraction involves mappings from multiple vSDN switches
to one physical SDN switch~\cite{gho2014tow,she2010can}. 
Before we go on to give general background on the two main hypervisor functions,
namely the abstraction (virtualization) of network attributes in
Section~\ref{sec:abstraction} and the isolation of the
vSDNs in Section~\ref{sec:isolation}, we briefly review
the literature on hypervisors.
Hypervisors for virtual computing systems have been surveyed
in~\cite{des2013hyp,kiv2007kvm,mal2014sur,ros2005vir},
while hypervisor vulnerabilities in cloud computing
have been surveyed in~\cite{per2013char}.
However, to the best of our knowledge there has been no prior detailed survey
of hypervisors for virtualizing SDN networks into vSDNs.
We comprehensively survey SDN hypervisors in this article.

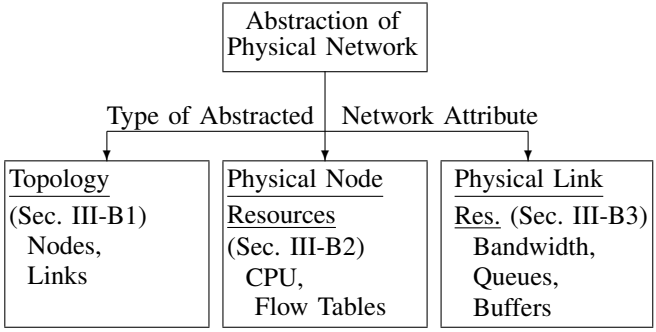
\begin{figure}[t]
\setlength{\unitlength}{1mm}
\centering
\begin{picture}(85,52)
\put(29,43){\makebox(0,0)[lt]{\framebox(27,9)}}
\put(29.5,42){\makebox(0,0)[lt]{\shortstack[c]{Abstraction of \\
                    Physical  Network }}}
\put(42.5,34){\vector(0,-1){12}}
\put(13.5,27.75){\makebox(0,0)[l]
  {\shortstack[c]{Type of Abstracted\ \ \ Network Attribute}}}
\put(13.5,26){\line(1,0){56}} \put(13.5,26){\vector(0,-1){4}}
\put(69.5,26){\vector(0,-1){4}}
\put(0,22){\makebox(0,0)[lt]{\framebox(27,22)}}
\put(0.5,21){\makebox(0,0)[lt]{\shortstack[l]{\underline{Topology} \\
  (Sec.~\ref{topabs:sec}) \\
 \ \ Nodes, \\
 \ \ Links }}}
\put(29,22){\makebox(0,0)[lt]{\framebox(27,22)}}
\put(29.5,21){\makebox(0,0)[lt]{\shortstack[l]{\underline{Physical Node} \\
 \underline{Resources}\\
  (Sec.~\ref{absphyn:sec})  \\
    \ \ CPU, \\\
    \ \ Flow Tables }}}
\put(58,22){\makebox(0,0)[lt]{\framebox(28,22)}}
\put(58.5,21){\makebox(0,0)[lt]{
       \shortstack[l]{\underline{Physical Link} \\
        \underline{Res.} (Sec.~\ref{absphyl:sec})\\
    \ \  Bandwidth,\\
    \ \   Queues, \\
    \ \   Buffers }}}
\end{picture}
\caption{The hypervisor abstracts three types of physical SDN network
attributes, namely topology, node resources, and link resources, with
different levels of virtualization, e.g., a given physical network topology is
abstracted to more or less virtual nodes or links.}
    \label{fig:lev_virt}
\end{figure}
\subsection{Network Attribute Virtualization}  \label{sec:abstraction}
The term ``abstraction'' can generally be defined as~\cite{dict02}:
the act of considering something as a general quality or characteristic,
apart from concrete realities, specific objects, or actual instances.
In the context of an SDN network,
a hypervisor abstracts the specific characteristic details
(attributes) of the underlying physical SDN network.
The abstraction (simplified representation) of the SDN
network is communicated by the hypervisor to the controllers of
the virtual tenants, i.e., the virtual SDN controllers.
We refer to the degree of simplification (abstraction) of the
network representation as the \textit{level of virtualization}.

Three types of attributes of the physical SDN network,
namely topology, physical node resources, and physical link resources,
are commonly considered in SDN network abstraction.
For each type of network attribute, we consider
different levels (degrees, granularities) of virtualization,
as summarized in Fig~\ref{fig:lev_virt}.

\subsubsection{Topology Abstraction} \label{topabs:sec}
For the network attribute topology,
virtual nodes and virtual links define the level of virtualization.
The hypervisor can abstract an end-to-end network path traversing
multiple physical links as one end-to-end virtual 
link~\cite{Chowdhury2008,cas2010vir}.
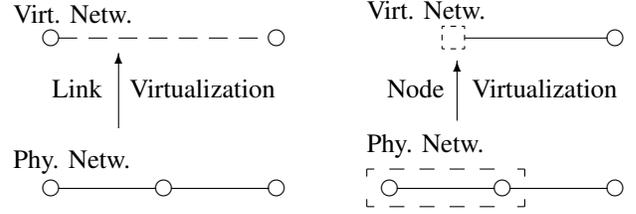
\begin{figure}[t]
\setlength{\unitlength}{1mm}
\centering
\begin{picture}(85,30)
\put(0,22){\makebox(0,0)[lb]{\shortstack[c]{Virt.  Netw.}}}
\put(5,20){\circle{2}} \multiput(6,20)(4,0){7}{\line(1,0){2.5}}
\put(35,20){\circle{2}}
\put(20,12){\makebox(0,0)[b]{\shortstack[c]{Link \ \ Virtualization}}}
\put(14,8){\vector(0,1){10}}
\put(0,2){\makebox(0,0)[lb]{\shortstack[c]{Phy.  Netw.}}}
\put(5,0){\circle{2}} \put(6,0){\line(1,0){13}} \put(20,0){\circle{2}}
\put(21,0){\line(1,0){13}}   \put(35,0){\circle{2}}
\put(65,12){\makebox(0,0)[b]{\shortstack[c]{Node \ \ Virtualization}}}
\put(59,9){\vector(0,1){8}}
\put(47,22.5){\makebox(0,0)[lb]{\shortstack[c]{Virt.  Netw.}}}
\put(57,18.5){\makebox(0,0)[lb]{\dashbox{0.75}(3,3)}}
\put(60,20){\line(1,0){19}}   \put(80,20){\circle{2}}
\put(47,4){\makebox(0,0)[lb]{\shortstack[c]{Phy.  Netw.}}}
\put(50,0){\circle{2}} \put(51,0){\line(1,0){13}} \put(65,0){\circle{2}}
\put(66,0){\line(1,0){13}}   \put(80,0){\circle{2}}
\put(47,-2.5){\makebox(0,0)[lb]{\dashbox{2}(21,5)}}
\end{picture}
\caption{Illustration of link and node virtualization:
For link virtualization, the two physical network links and
intermediate network node are virtualized to a single virtual link
(drawn as a dashed line).
For node virtualization, the left two nodes and the physical link
connecting the two nodes are virtualized to a single virtual node
(drawn as a dashed box).
}
    \label{fig:virt}
\end{figure}
A simple example of link virtualization where two physical links
(and an intermediate node) are virtualized to a single virtual link
is illustrated in the left part of Fig.~\ref{fig:virt}.
For another example of link virtualization
consider the left middle node in Fig.~\ref{concepts:fig}(b),
which is connected via two links and the upper left node with the node in the
top middle. These two links and the intermediate (upper left corner) node are
abstracted to a single virtual link (drawn as a dashed line)
in vSDN~1 in the left part of Fig.~\ref{concepts:fig}(c).
Similarly, the physical link from the left middle node in
Fig.~\ref{concepts:fig}(b) via the bottom right corner node to
the node in the bottom middle is abstracted to a single
virtual link (dashed line) in the left part of Fig.~\ref{concepts:fig}(c).

Alternatively, the hypervisor can abstract multiple physical nodes to
a single virtual node~\cite{gho2014tow}.
In the example illustrated in the right
part of Fig.~\ref{fig:virt}, two physical nodes and
their connecting physical link are virtualized to a single
virtualized node (drawn as a dashed box).

The least (lowest) level of virtualization of the topology
represents the physical nodes and physical links in an
identical virtual topology, i.e., the abstraction is a transparent
1-to-1 mapping in that the physical topology is \textit{not}
modified to obtain the virtual topology.
The highest level of virtualization abstracts
the entire physical network topology as a single virtual link or node.
Generally, there is a range of levels of virtualization between the
outlined lowest and highest levels of virtualization 
link~\cite{Chowdhury2008}.
For instance, a complex physical network can be abstracted to a
simpler virtual topology with fewer virtual nodes and links,
as illustrated in the view of vSDN~1 controller in Fig.~\ref{concepts:fig}(c)
of the actual physical network in Fig.~\ref{concepts:fig}(b).
Specifically, the vSDN~1 controller in Fig.~\ref{concepts:fig}(c)
``sees'' only a vSDN~1 consisting of four nodes interconnected by five links,
whereas the underlying physical network illustrated in
Fig.~\ref{concepts:fig}(b) has six nodes interconnected by seven links.
This abstraction is sometimes referred to as 1-to-$N$ mapping 
or as one-to-many mapping as a single virtual node or link is mapped to
several physical nodes or links~\cite{gho2014tow}. 
In particular, the mapping of one virtual switch to multiple physical switches
is also sometimes referred to as the ``big switch'' 
abstraction~\cite{mon2013com}.

\subsubsection{Abstraction of Physical Node Resources} \label{absphyn:sec}
For the network attribute physical node resources,
mainly the CPU resources and the flow table resources
of an SDN switch node define the level of virtualization~\cite{Chowdhury2008}.
Depending on the available level of CPU hardware information,
the CPU resources may be represented by a wide range of CPU
characterizations, e.g., number of assigned CPU cores or
percentage of CPU capacity in terms of the capacity of a single core,
e.g., for a CPU with two physical cores, 150~\% of the capacity corresponds
to one and a half cores.
The flow table abstraction may similarly involve a wide range of
resources related to flow table processing, e.g., the number of flow tables
or the number of ternary content-addressable memories
(TCAMs)~\cite{pag2006con,pan2002red} for flow table processing.
The abstraction of the SDN switch physical resources can be
beneficial for the tenants' vSDN controllers.

\subsubsection{Abstraction of Physical Link Resources} \label{absphyl:sec}
For the network attribute physical link resources,
the bandwidth (transmission bit rate, link capacity) as well as
the available link queues and link buffers
define the level of virtualization~\cite{Chowdhury2008}.
\begin{figure}[t]
\setlength{\unitlength}{1mm}
\centering
\begin{picture}(85,50)
\put(29,43){\makebox(0,0)[lt]{\framebox(27,9)}}
\put(29.5,42){\makebox(0,0)[lt]{\shortstack[c]{Isolation of Virt.\\
          SDN Networks}}}
\put(42.5,34){\vector(0,-1){12}}
\put(13.5,27.75){\makebox(0,0)[l]
  {\shortstack[c]{Type of Isolated\ \ \ \ \ \ Network Attribute}}}
\put(13.5,26){\line(1,0){56}} \put(13.5,26){\vector(0,-1){4}}
\put(69.5,26){\vector(0,-1){4}}
\put(0,22){\makebox(0,0)[lt]{\framebox(27,25)}}
\put(0.5,21){\makebox(0,0)[lt]{\shortstack[l]{\underline{Control Plane} \\
 \underline{Hypervisor} \\
  (Sec.~\ref{cpi:sec}) \\
   \ \ \ \   Instances,\\\
  \ \ \ \    Network}}}
\put(29,22){\makebox(0,0)[lt]{\framebox(27,22)}}
\put(29.5,21){\makebox(0,0)[lt]{\shortstack[l]{\underline{Data Plane} \\
 \underline{Physical Network}\\
(Sec.~\ref{dpi:sec}) \\
    \ \ \ \  Nodes, \\\
    \ \ \ \   Links}}}
\put(58,22){\makebox(0,0)[lt]{\framebox(28,22)}}
\put(58.5,21){\makebox(0,0)[lt]{
       \shortstack[l]{\underline{Virtual SDN} \\
        \underline{Netw. Addressing}\\
        (Sec.~\ref{vsdnai:sec}) \\
        \ \ \ \   Address \\
         \ \ \ \  Domain }}}
\end{picture}
\caption{The hypervisor isolates three main attributes of virtual
SDN networks, namely control plane, data plane, and vSDN addressing.}
\label{fig:iso_tree}
\end{figure}
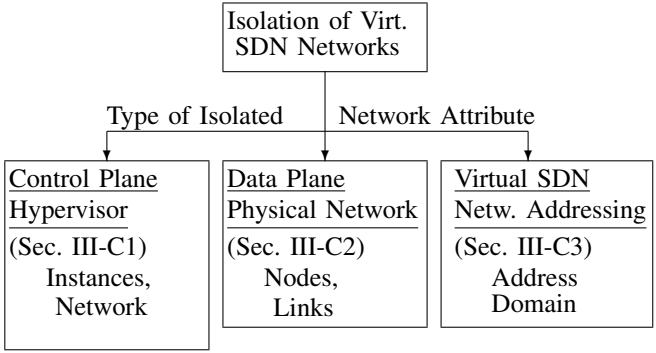

\subsection{Isolation Attributes}    \label{sec:isolation}
The hypervisor should provide isolated slices for the vSDNs
sharing a physical SDN network.
As summarized in Fig.~\ref{fig:iso_tree}, the isolation should cover
the SDN control plane and the SDN data plane.
In addition, the vSDN addressing needs to be isolated,
i.e., each vSDN should have unique flow identifiers.

\subsubsection{Control Plane Isolation} \label{cpi:sec}
SDN decouples the data plane from the control plane.
However, the performance of the control plane impacts the data plane
performance~\cite{too2010hyp,ToGGC12}.
Thus, isolation of the tenants' control planes is needed.
Each vSDN controller should have the impression of controlling its
own vSDN without interference from other vSDNs (or their controllers).

The control plane performance is influenced by the resources
available for the hypervisor instances on
the hosting platforms, e.g., commodity servers or
NEs~\cite{Epfl-report-,rot2012ofl}.
Computational resources, e.g., CPU resources, can affect the performance of the
hypervisor packet processing and translation, while storage resources
limit control plane buffering.
Further, the network resources of the links/paths forming
the hypervisor layer, e.g., link/path data rates and the buffering capacities
need to be isolated to provide control plane isolation.

\subsubsection{Data Plane Isolation}  \label{dpi:sec}
Regarding the data plane, physical node resources mainly relate to
network element CPU and flow tables.
Isolation of the physical link resources relates to the
link transmission bit rate.
More specifically, the performance of the data plane physical nodes (switches)
depends on their processing
capabilities, e.g., their CPU capacity and their hardware
accelerators.  Under different work loads, the utilization of these
resources may vary, leading to varying performance, i.e., changing
delay characteristics for packet processing actions.
Switch resources should be assigned (reserved) for a given vSDN
to avoid performance degradation.
Isolation mechanisms should prevent cross-effects between vSDNs,
e.g., that one vSDN over-utilizes its assigned switch element
resources and starves another vSDN of its resources.

Besides packet processing capabilities, an intrinsic characteristic of
an SDN switch is the capacity for storing and matching flow rules.
The OF protocol stores rules in flow tables.
Current OF-enabled switches often use fast TCAMs for
storing and matching rules.  For proper isolation,
specific amounts of TCAM capacity should be assigned to the
vSDNs.

Hypervisors should provide physical link transmission bit rate
isolation as a prerequisite towards providing Quality-of-Service (QoS).
For instance, the transmission rate may need to be isolated to achieve
low latency.
Thus, the hypervisor should be able to assign a specific amount of
transmission rate to each tenant. Resource allocation and packet scheduling
mechanisms, such as Weighted Fair Queuing
(WFQ)~\cite{ben1997hie,lee2014des,sti1998eff},
may allow assigning prescribed amounts of resources.
In order to impact the end-to-end latency, mechanisms that assign the
available buffer space on a per-link or per-queue basis are needed as well.
Furthermore, the properties of the queue operation, such as
First-In-First-Out (FIFO) queueing, may impact specific isolation properties.

\subsubsection{vSDN Addressing Isolation}  \label{vsdnai:sec}
As SDN follows a match and action paradigm (see Section~\ref{bg_sdn:sec}),
flows from different vSDNs must be uniquely addressed (identified).
An addressing solution must guarantee
that forwarding decisions of tenants do not conflict with each other.
The addressing should also maximize the flowspace, i.e., the match set,
that the vSDN controllers can use.
One approach is to split the flowspace,
i.e., the match set, and provide the tenants with non-overlapping flowspaces.
If a vSDN controller uses an OF version, e.g., OF v1.0, lower than the
hypervisor, e.g., OF v1.3, then the hypervisor could use the extra fields
in the newer OF version for tenant identification. This way such a controller
can use the full flowspace of its earlier OF version for the operation of
its slice.
However, if the vSDN controller implements the same or a newer OF version
than the hypervisor, then the vSDN controller cannot use the full flowspace.

Another addressing approach is to use fields outside of the OF matching
flowspace, i.e., fields not defined in the OF specifications,
for the unique addressing of vSDNs. In this case, independent of the
implemented OF version of the vSDN controllers, the full flowspace
can be offered to the tenants.

\section{Classification Structure of Hypervisor Survey}
\label{class:sec}
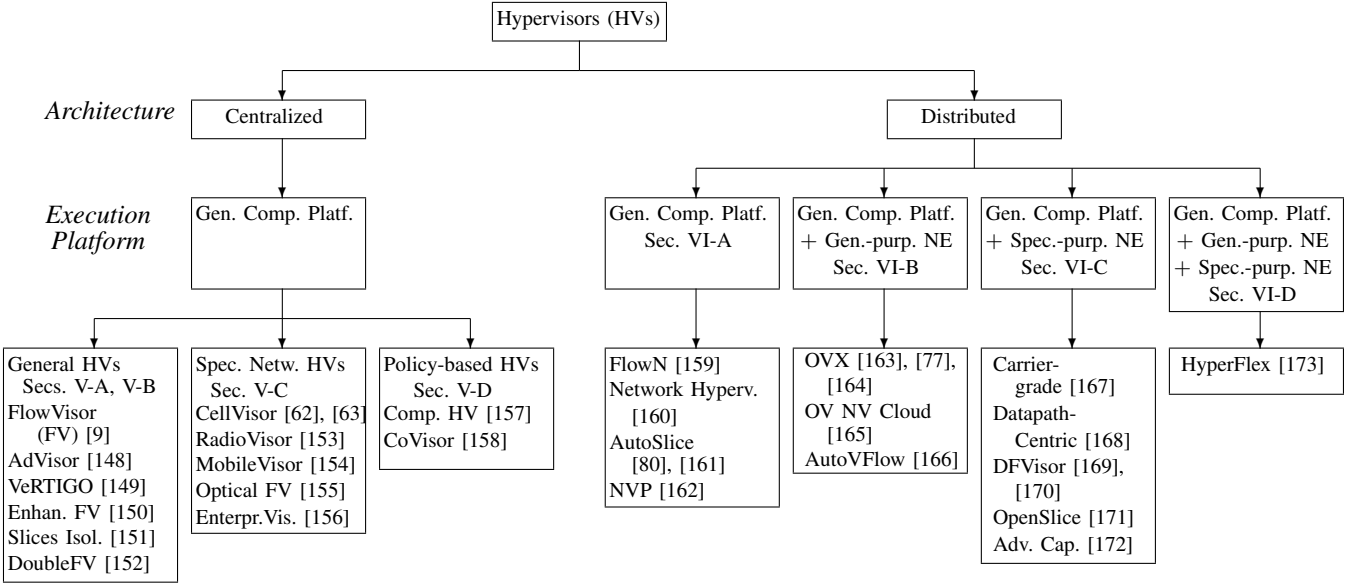
\begin{figure*}[t]
    \centering
\footnotesize
\setlength{\unitlength}{1mm}
\centering
\begin{picture}(180,76)
\put(65,76){\makebox(0,0)[lt]{\framebox(23,5)}}
\put(65.5,75){\makebox(0,0)[lt]{\shortstack[c]{Hypervisors (HVs)}}}
\put(76.5,71){\line(0,-1){4}}
\put(76.5,67){\line(-1,0){39.5}}
\put(37,67){\vector(0,-1){4}}
\put(76.5,67){\line(1,0){52.5}}
\put(129,67){\vector(0,-1){4}}
\put(5.5,63){\makebox(0,0)[lt]
  {\shortstack[c]{{\normalsize \textit{Architecture}}}}}
\put(25,63){\makebox(0,0)[lt]{\framebox(23,5)}}
\put(25.5,62){\makebox(0,0)[lt]{\shortstack[c]{\ \ \ \ Centralized}}}
\put(37,58){\vector(0,-1){8}}
\put(117.5,63){\makebox(0,0)[lt]{\framebox(23,5)}}
\put(118,62){\makebox(0,0)[lt]{\shortstack[c]{\ \ \ \ Distributed}}}
\put(129,58){\line(0,-1){4}}
\put(92,54){\line(1,0){75}}
\put(92,54){\vector(0,-1){4}}
\put(117,54){\vector(0,-1){4}}
\put(142,54){\vector(0,-1){4}}
\put(167,54){\vector(0,-1){4}}
\put(5.5,49){\makebox(0,0)[lt]
  {\shortstack[c]{ {\normalsize \textit{Execution}} \\
                  {\normalsize \textit{Platform}}}}}
\put(25,50){\makebox(0,0)[lt]{\framebox(23,12)}}
\put(25.5,49){\makebox(0,0)[lt]{\shortstack[c]{Gen. Comp. Platf. }}}
\put(37,38){\vector(0,-1){8}}
\put(12,34){\line(1,0){50}}
\put(12,34){\vector(0,-1){4}}
\put(62,34){\vector(0,-1){4}}
\put(80,50){\makebox(0,0)[lt]{\framebox(23,12)}}
\put(80.5,49){\makebox(0,0)[lt]{\shortstack[c]{Gen. Comp. Platf.\\
    Sec.~\ref{gcpdishyp:sec}}}}
\put(92,38){\vector(0,-1){8}}
\put(105,50){\makebox(0,0)[lt]{\framebox(23,12)}}
\put(105.5,49){\makebox(0,0)[lt]{\shortstack[c]{Gen. Comp. Platf.  \\
    $+$ Gen.-purp. NE\\
    Sec.~\ref{gcpgpne:sec}}}}
\put(117,38){\vector(0,-1){8}}
\put(130,50){\makebox(0,0)[lt]{\framebox(23,12)}}
\put(130.5,49){\makebox(0,0)[lt]{\shortstack[c]{Gen. Comp. Platf.  \\
    $+$ Spec.-purp. NE\\
    Sec.~\ref{gcpspne:sec}}}}
\put(142,38){\vector(0,-1){8}}
\put(155,50){\makebox(0,0)[lt]{\framebox(23,15)}}
\put(155.5,49){\makebox(0,0)[lt]{\shortstack[c]{Gen. Comp. Platf.  \\
    $+$ Gen.-purp. NE  \\
    $+$ Spec.-purp. NE\\
    Sec.~\ref{gcpgspne:sec}  }}}
\put(167,35){\vector(0,-1){5}}
\put(0,30){\makebox(0,0)[lt]{\framebox(23,31)}}
\put(0.5,29){\makebox(0,0)[lt]{\shortstack[l]{General HVs \\
 \ \ Secs.~\ref{FlowVisor:sec},~\ref{cenotherhyp:sec} \\
    FlowVisor \\
   \ \ \ \ (FV)~\cite{Sherwood2009} \\
         AdVisor~\cite{Salvadori2011} \\
     VeRTIGO~\cite{Corin2012} \\
      Enhan.~FV~\cite{Min2012} \\
     Slices Isol.~\cite{El-Azzab2011} \\
    DoubleFV~\cite{Yin2013}   }}}
\put(25,30){\makebox(0,0)[lt]{\framebox(23,24.5)}}
\put(25.5,29){\makebox(0,0)[lt]{\shortstack[l]{Spec. Netw. HVs\\
    \ \ Sec.~\ref{specenhyp:sec} \\
     CellVisor~\cite{li2012tow,Li} \\
    RadioVisor~\cite{Gudipati2014} \\
    MobileVisor~\cite{Nguyen2014} \\ Optical FV~\cite{Azodolmolky2012} \\
   Enterpr.Vis.~\cite{Chen2014a} }}}
\put(50,30){\makebox(0,0)[lt]{\framebox(23,16)}}
\put(50.5,29){\makebox(0,0)[lt]{\shortstack[l]{Policy-based HVs\\
                 \ \ \ \ Sec.~\ref{polcenhyp:sec} \\
    Comp.~HV~\cite{Rexford2014} \\
     CoVisor~\cite{Jin2015}    }}}
\put(80,30){\makebox(0,0)[lt]{\framebox(23,21)}}
\put(80.5,29){\makebox(0,0)[lt]{\shortstack[l]{FlowN~\cite{Drutskoy2013b} \\
   Network Hyperv. \\
 \ \ \ \cite{hua2013net} \\
    AutoSlice \\
    \ \ \ \cite{Bozakov2012,Bozakov2013} \\
    NVP~\cite{Koponen2014} }}}
\put(105,30){\makebox(0,0)[lt]{\framebox(23,16.5)}}
\put(105.5,29.5){\makebox(0,0)[lt]{
    \shortstack[l]{ OVX~\cite{Al-Shabibi2014},~\cite{Berde2014}, \\
                  \ \ \ \cite{Al-shabibi2014b}\\
    OV NV Cloud \\ \ \ \ \cite{Matias2011} \\
   AutoVFlow~\cite{Yamanaka2014} }}}
\put(130,30){\makebox(0,0)[lt]{\framebox(23,28.5)}}
\put(130.5,29){\makebox(0,0)[lt]{
    \shortstack[l]{Carrier-\\
    \ \ \ grade~\cite{Devlic2012a}\\
    Datapath- \\
    \ \ \ Centric~\cite{DoriguzziCorin2014} \\
    DFVisor~\cite{Liao2014},\\
     \ \ \ \cite{Liao2015}\\
   OpenSlice~\cite{liu2013os} \\
   Adv. Cap.~\cite{Sonkoly2012b} }}}
\put(155,30){\makebox(0,0)[lt]{\framebox(23,6)}}
\put(155.5,29){\makebox(0,0)[lt]{
    \shortstack[l]{HyperFlex~\cite{ble2015hyp}}}}
\end{picture}
    \caption{Hypervisor Classification: Our top-level classification
criterion is the hypervisor architecture (centralized or distributed).
Our second-level classification is according to the hypervisor
execution platform (general-purpose compute platform only or combination
of general-purpose compute platform and general- or special-purpose network
elements). For the large set of centralized compute-platform hypervisors,
we further distinguish hypervisors for special network types
(wireless, optical, or enterprise network) and policy-based hypervisors.}
    \label{fig:class}
\end{figure*}

In this section, we introduce our hypervisor classification,
which is mainly based on the architecture and the execution platform, as
illustrated in Fig.~\ref{fig:class}.
We first classify SDN hypervisors according to their
architecture into centralized and distributed hypervisors.
We then sub-classify the hypervisors according to their
execution platform into hypervisors
running exclusively on general-purpose compute platforms as well as hypervisors
running on general-purpose computing platforms in
combination with general- or/and special-purpose NEs.
We proceed to explain these classification categories in detail.
\normalsize

\subsection{Type of Architecture}
A hypervisor has a centralized architecture
if it consists of a single central entity.
This single central entity controls multiple network elements (NEs), i.e.,
OF switches, in the physical network infrastructure.
Also, the single central entity serves potentially multiple
tenant controllers.
Throughout our classification, we classify hypervisors that do not require
the distribution of their hypervisor functions as centralized.
We also classify hypervisors as centralized, when no detailed distribution
mechanisms for the hypervisor functions have been provided.
We sub-classify the centralized hypervisors into hypervisors for
general networks, hypervisors for special network types
(e.g., optical or wireless networks), and policy-based hypervisors.
Policy-based hypervisors allow multiple network applications through
different vSDN controllers,
e.g., $\mbox{App}_{11}$ through vSDN 1 Controller and
$\mbox{App}_{21}$ through vSDN 2 Controller in
Fig.~\ref{concepts:fig}(b), to ``logically''
operate on the same traffic flow in the
underlying physical SDN network.
The policy-based hypervisors compose the OF rules of the two
controllers to achieve this joint operation on the same traffic flow,
as detailed in Section~\ref{polcenhyp:sec}.

We classify an SDN hypervisor as a distributed hypervisor if the
virtualization functions can run logically separated from each other.
A distributed hypervisor appears logically as consisting of
a single entity (similar to a centralized hypervisor); however,
a distributed hypervisor consists of several distributed functions.
A distributed hypervisor may decouple management functions from translation
functions or isolation functions. However, the hypervisor
functions may depend on each other. For instance, in order to protect the
translation functions from over-utilization, the isolation functions should
first process the control traffic.
Accordingly, the hypervisor needs to provide mechanisms for orchestrating
and managing the functions, so as to guarantee the valid operation of
dependent functions. These orchestration and management mechanisms could be
implemented and run in a centralized or in a distributed manner.

\subsection{Hypervisor Execution Platform}
We define the hypervisor execution platform to characterize
(hardware) infrastructure (components) employed for
implementing (executing) a hypervisor.
Hypervisors are commonly implemented through
software programs (and sometimes employ specialized NEs).
The existing centralized hypervisors are implemented through
software programs that run on general-purpose
compute platforms, e.g., commodity compute servers and personal computers (PCs),
henceforth referred to as ``compute platforms'' for brevity.

Distributed hypervisors may employ compute platforms
in conjunction with general-purpose NEs or/and special-purpose
NEs, as illustrated in Fig.~\ref{fig:class}.
We define a general-purpose NE to be a commodity off-the-shelf switch
without any specialized hypervisor extensions.
We define a special-purpose NE to be a customized switch that has
been augmented with specialized hypervisor functionalities
or extensions to the OF specification, such as the capability to
match on labels outside the OF specification.

We briefly summarize the pros and cons of the different
execution platforms as follows.
Software implementations offer ``portability'' and ease of operation on
a wide variety of general-purpose (commodity) compute platforms.
A hypervisor implementation
on a general-purpose (commodity) NE can utilize the existing
hardware-integrated NE processing
capabilities, which can offer higher performance compared to general-purpose
compute platforms. However, general-purpose NEs may lack some of the
hypervisor functions required for creating vSDNs.
Hence commodity NEs can impose limitations.
A special-purpose NE offers high hardware performance and covers the set of
hypervisor functions. However replacing commodity NEs in a network
with special-purpose NEs can be prohibitive
due to the additional cost and equipment migration effort.

As illustrated in Fig.~\ref{fig:class}, we sub-classify the distributed
hypervisors into hypervisors designed to execute on compute platforms,
hypervisors executing on compute platforms in
conjunction with \textit{general-purpose} NEs,
 hypervisors executing on compute platforms in
conjunction with \textit{special-purpose} NEs,
as well as  hypervisors executing on compute platforms in
conjunction with a mix of general- and special-purpose NEs.

\section{Centralized Hypervisors}
\label{central:sec}
In this section, we comprehensively survey the existing hypervisors (HVs)
with a centralized architecture.
All existing centralized hypervisors are executed on a central
general-purpose computing platform.
FlowVisor~\cite{Sherwood2009} was a seminal hypervisor for virtualizing
SDN networks. We therefore dedicate a separate subsection,
namely Section~\ref{FlowVisor:sec}, to a detailed overview of FlowVisor.
We survey the other hypervisors for general networks in
Section~\ref{cenotherhyp:sec}.
We cover hypervisors for special network types in Section~\ref{specenhyp:sec},
while policy-based hypervisors are covered in Section~\ref{polcenhyp:sec}.

\subsection{FlowVisor}
\label{FlowVisor:sec}
\subsubsection{General Overview}
FlowVisor~\cite{Sherwood2009} was the first hypervisor for virtualizing and
sharing software defined networks based on the OF protocol.
The main motivation for the development of FlowVisor
was to provide a hardware abstraction layer that facilitates
innovation above and below the virtualization layer.
In general, a main FlowVisor goal is to run production and
experimental networks on the same physical SDN networking hardware.
Thus, FlowVisor
particularly focuses on mechanisms to isolate experimental network
traffic from production network traffic.
Additionally, general
design goals of FlowVisor are to work in a transparent manner, i.e.,
without affecting the hosted virtual networks, as well as to provide
extensible, modular, and flexible definitions of network slices.

\subsubsection{Architecture} FlowVisor is a pure software implementation.
It can be deployed on general-purpose (commodity) computing servers
running commodity operating systems.
In order to control and manage access to the slices, i.e., physical hardware,
FlowVisor sits between the tenant's controllers and the
physical (hardware) SDN network switches, i.e., at the position of the
box marked ``Hypervisor'' in Fig.~\ref{concepts:fig}.
Thus, FlowVisor controls the views that the tenants' controllers
have of the SDN switches.
FlowVisor also controls the access of the tenants' controllers to the
switches.
\begin{figure*}[ht]
    \centering
  \includegraphics[width=0.85\textwidth]{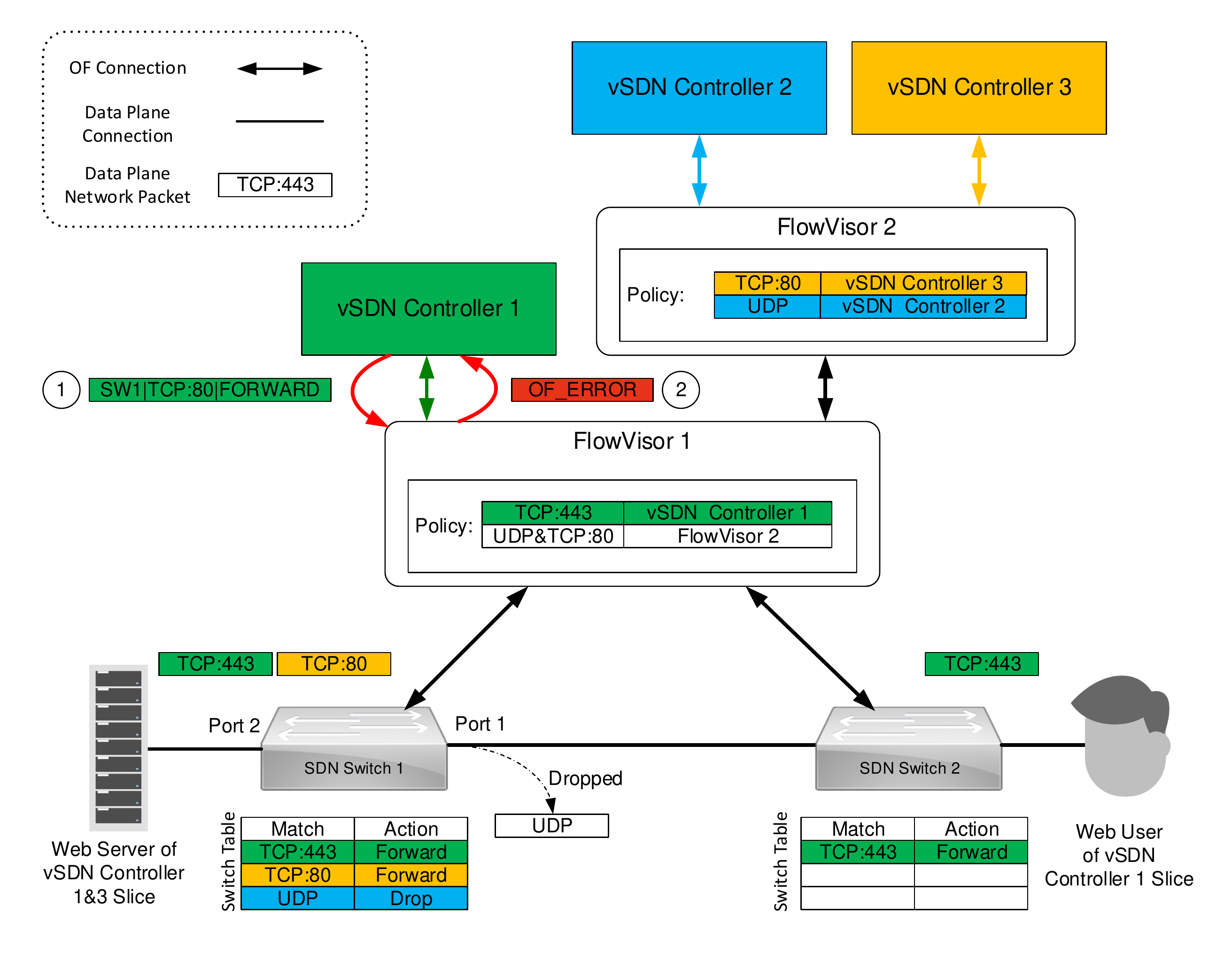}
    \caption{Illustrative example of SDN network virtualization with
      FlowVisor: FlowVisor~1 creates a vSDN that vSDN Controller~1
    uses to control HTTPS traffic with TCP port number 443 on
   SDN switches~1 and~2. FlowVisor~2 is nested on top of FlowVisor~1
   and controls only SDN Switch~1.
   FlowVisor~2 lets vSDN Controller~3 control the HTTP traffic with TCP port
  number 80 on SDN Switch~1 (top-priority rule in FlowVisor 2,
   second-highest priority rule in SDN Switch~1) and
lets vSDN Controller~2 drop all UDP traffic.}
    \label{fig:flowvisor}
\end{figure*}

FlowVisor introduces the term \textit{flowspace} for a sub-space of
the header fields space~\cite{OF15} of an OF-based network.
FlowVisor allocates each vSDN (tenant) its own flowspace, i.e., its own
sub-space of the OF header fields space and ensures that the
flowspaces of distinct vSDNs do not overlap.
The vSDN controller of a given tenant operates on its flowspace, i.e.,
 its sub-space of the OF header fields space.
The flowspace concept is illustrated for an
OF-based example network shared by three tenants in Fig.~\ref{fig:flowvisor}.
The policy (prioritized list of forwarding rules) of vSDN Controller 1
specifies the control of all HTTPS traffic (with highest priority);
thus, vSDN Controller~1 controls all packets whose TCP port header
field matches the value~443.
The policy of a second vSDN tenant of FlowVisor~1, which is a nested
instance of a hypervisor, namely FlowVisor~2,
specifies to control all HTTP (TCP port value 80) and UDP traffic.
FlowVisor~1 defines policies through the
matching of OF header fields to guarantee that the
virtual slices do not interfere with each other, i.e.,
the virtual slices are isolated from each other.

\subsubsection{Abstraction and Isolation Features}  \label{absisoftfv:sec}
FlowVisor provides the vSDNs with
bandwidth isolation, topology isolation, switch CPU isolation,
flowspace isolation, isolation of the flow entries,
and isolation of the OF control channel (on the D-CPI).
The FlowVisor version examined in~\cite{Sherwood2009}
maps the packets of a given slice to a prescribed
Virtual Local Area Network (VLAN) Priority Code Point (PCP).
The 3-bit VLAN PCP allows for the mapping to eight distinct priority
queues~\cite{sof2009sur}.
More advanced
bandwidth isolation (allocation) and scheduling mechanisms are evaluated in
research that extends FlowVisor, such as Enhanced
FlowVisor~\cite{Min2012}, see Section~\ref{EnFV:sec}.

For topology isolation, only the physical
resources, i.e., the ports and switches, that are part of a slice are
shown to the respective tenant controller.
FlowVisor achieves the topology isolation by acting as a proxy
between the physical resources and the tenants' controllers.
Specifically, FlowVisor edits OF messages to only report
the physical resources of a given slice to the corresponding
tenant controller.
Fig.~\ref{fig:flowvisor} illustrates the topology abstraction.
In order to provide secure HTTPS connections to its applications,
vSDN Controller~1 controls a slice spanning SDN Switches~1 and~2.
In contrast, since the tenants of FlowVisor~2 have only traffic traversing
SDN Switch~1, FlowVisor~2 sees only SDN Switch 1.
As illustrated in Fig.~\ref{fig:flowvisor}, vSDN Controller~1 has
installed flow rules on both switches, whereas vSDN Controllers~2 and~3
have only rules installed on SDN Switch~1.

The processing of OF messages can overload the
central processing unit (CPU) in a physical SDN switch, rendering the
switch unusable for effective networking.
In order to ensure that each slice is effectively supported by the
switch CPU, FlowVisor limits the rates of the different types of
OF messages exchanged between physical switches and the corresponding
tenant controllers.

The flowspaces of distinct vSDNs are not
allowed to overlap. If one tenant controller tries to set a rule that
affects traffic outside its slice, FlowVisor rewrites
such a rule to the tenant's slice. If a rule cannot be rewritten, then
FlowVisor sends an OF error message.
Fig.~\ref{fig:flowvisor} illustrates the flowspace isolation.
When vSDN Controller~1 sends an
OF message to control HTTP traffic with TCP port field value~80, i.e.,
traffic that does not match TCP port~443, then
FlowVisor~1 responds with an OF error message.

Furthermore, in order to provide a simple means for defining varying
policies, FlowVisor instances can run on top of each other. This means
that they can be nested in order to provide different levels of
abstraction and policies.
As illustrated in Fig.~\ref{fig:flowvisor}, FlowVisor~2 runs on top of
FlowVisor~1. FlowVisor~1 splits the OF header fields space
between vSDN Controller~1 and FlowVisor~2.
In the illustrated example, vSDN Controller~1 controls
HTTPS traffic. FlowVisor~2, in turn serves vSDN Controller~2, which
implements a simple firewall that drops all UDP packets,
and vSDN Controller~3, which implements an HTTP traffic controller.
The forwarding rules for HTTPS (TCP port 443) packet traffic and HTTP
(TCP port 80) packet traffic are listed higher, i.e., have higher priority,
than the drop rule of vSDN Controller~2.
Thus, SDN Switch~1 forwards HTTPS and HTTP traffic, but drops other traffic,
e.g., UDP datagrams.

SDN switches typically store OF flow entries (also sometimes referred to as
OF rules), in a limited amount of TCAM memory, the so-called flow table memory.
FlowVisor assigns each tenant a part of the flow table memory in each
SDN switch.
In order to provide isolation of flow entries, each switch keeps track of
the number of flow entries inserted by a tenant controller.
If a tenant exceeds a prescribed limit of flow entries, then FlowVisor
replies with a message indicating that the flow table of the switch is full.

The abstraction and isolation features reviewed so far relate to the
physical resources of the SDN network. However, for effective
virtual SDN networking, the data-controller plane interface (D-CPI)
(see Section~\ref{bg_sdn:sec}) should also be isolated.
FlowVisor rewrites the OF transaction identifier to ensure that
the different vSDNs utilize distinct transaction identifiers.
Similarly, controller buffer accesses and status messages are
modified by FlowVisor to create isolated OF control slices.

FlowVisor has been extended with an intermediate control plane slicing layer
that contains a Flowspace Slicing Policy (FSP) engine~\cite{Argy2015}.  
The FSP engine adds three control plane slicing methods:
domain-wide slicing, switch-wide slicing, and port-wide slicing. The
three slicing methods differ in their proposed granularity of
flowspace slicing. With the FSP engine, tenants can request abstracted
virtual networks, e.g., end-to-end paths only. According to the
demanded slicing policy, FSP translates the requests into multiple
isolated flowspaces. The concrete flowspaces are realized via an
additional proxy, e.g., FlowVisor. The three slicing methods are
evaluated and compared in terms of acceptance ratio, required hypervisor
memory, required switch flow table size, and additional control plane
latency added by the hypervisor. 
Specifying virtual networks more explicitly (with finer granularity), 
i.e., matching on longer header information, the port-wide
	slicing can accept the most virtual network demands while requiring
	the most resources.

\subsubsection{Evaluation Results}
The FlowVisor evaluations in~\cite{Sherwood2009}
cover the overhead as well as isolation of transmission (bit rate) rate, 
flowspace, and switch CPUs through measurements in a testbed.
FlowVisor sits as an additional
component between SDN controllers and SDN switches adding latency
overhead to the tenant control operations.
The measurements reported in~\cite{Sherwood2009}  indicate that
FlowVisor adds an average latency overhead of 16~ms for adding new flows and
0.48~ms for requesting port status.
These latency overheads can be reduced through optimized
request handling.

The bandwidth isolation experiments in~\cite{Sherwood2009}
let a vSDN (slice) sending a TCP flow compete with a vSDN
sending constant-bit-rate (CBR) traffic at the full
transmission bit rate of the shared physical SDN network.
The measurements indicate that without bandwidth isolation, the
TCP flow receives only about one percent of the link bit rate.
With bandwidth isolation that maps the TCP slice to a
specific class, the reported measurements indicate that
the TCP slice received close to the prescribed bit rate, while
the CBR slice uses up the rest of the bit rate.
Flowspace isolation is evaluated with a test suite that verifies
the correct isolation for 21 test cases, e.g., verifying that a
vSDN (slice) cannot manipulate traffic of another slice.
The  reported measurements for the OF message throttling mechanism indicate
that a malicious slice can fully saturate the experimental switch CPU with about
256 port status requests per second. The FlowVisor
switch CPU isolation, however, limits the switch CPU utilization to
less than 20~\%.

\subsection{General Hypervisors Building on FlowVisor}
\label{cenotherhyp:sec}
We proceed to survey the centralized hypervisors for general networks.
These general hypervisors
build on the concepts introduced by FlowVisor.
Thus, we focus on the extensions that each hypervisor provides
compared to FlowVisor.

\subsubsection{AdVisor}
AdVisor\cite{Salvadori2011} extends
FlowVisor in three directions.  First, it introduces an
improved abstraction mechanism that hides physical switches in virtual
topologies.
In order to achieve a more flexible topology abstraction, AdVisor
does not act as a transparent proxy between tenant (vSDN) controllers and
SDN switches.
Instead, AdVisor directly replies to the SDN switches.
More specifically, FlowVisor provides a transparent 1-to-1 mapping
with a detailed view of intermediate
physical links and nodes. That is, FlowVisor can present
1-to-1 mapped subsets of underlying topology to the vSDN controllers,
e.g., the subset of the topology illustrated in vSDN network 2
in the right part of  Fig.~\ref{concepts:fig}(c).
However, FlowVisor cannot ``abstract away'' intermediate physical nodes and
links, i.e., FlowVisor cannot create vSDN network~1 in the left part of
 Fig.~\ref{concepts:fig}(c).
In contrast, AdVisor extends the topology
abstraction mechanisms by hiding
intermediate physical nodes of a virtual path.  That is, AdVisor can show
only the endpoints of a virtual path to the tenants' controllers, and
thus create vSDN network~1 in the left part of
 Fig.~\ref{concepts:fig}(c).
When a physical SDN switch sends an OF message,
 AdVisor checks whether this message is from an endpoint of a virtual path.
If the switch is an endpoint, then the message is
forwarded to the tenant controller. Otherwise, i.e.,
if the switch has been ``abstracted away'', then AdVisor processes the
OF message and controls the forwarding of the traffic
independently from the tenant controller.

The second extension provided by AdVisor is the
sharing of the flowspace (sub-space of the OF header fields space)
by multiple (vSDNs) slices.
In order to support this sharing of the OF header fields space,
AdVisor defines the flowspace for the purpose of distinguishing vSDNs (slices)
to consist only of the bits of the OSI-layer 2. Specifically,
the AdVisor flowspace definition introduced so called \textit{slice\_tags},
which encompass \textit{VLAN id}, \textit{MPLS labels},
or IEEE802.1ad-based \textit{multiple VLAN tagging}~\cite{fou2009roa,sof2009sur,xia2000tra}.
This restricted definition of the AdVisor flowspace enables the
sharing of the remaining OF header fields among the slices.
However, labeling adds processing overhead to the NEs.
Furthermore, AdVisor is limited to NEs that provide labeling capabilities.
Restricting to VLAN ids limits the available number of slices.
When only using the VLAN id, the 4096 possible distinct VLAN ids
may not be enough for networks requiring many slices.
Large cloud providers have already on the order
of 1~Million customers today. If a large fraction of these customers were
to request their own virtual networks, the virtual networks
could not be distinguished.

\subsubsection{VeRTIGO}
VeRTIGO~\cite{Corin2012} takes the virtual network
abstraction of AdVisor~\cite{Salvadori2011} yet a step further.
VeRTIGO allows the vSDN controllers
to select the desired level of virtual network abstraction.
At the ``most detailed'' end of the abstraction spectrum,
VeRTIGO can provide the entire
set of assigned virtual resources, with full virtual network control.
At the other, ``least detailed'' end of the abstraction spectrum, VeRTIGO
abstracts the entire vSDN to a single abstract resource; whereby the
network operation is carried out by the hypervisor, while the tenant
focuses on services deployment on top.
While VeRTIGO gives high
flexibility in provisioning vSDNs, VeRTIGO has increased complexity.
In the evaluations reported in~\cite{Corin2012}, the average latencies
for new flow requests are increased roughly by 35~\% compared to FlowVisor.

\subsubsection{Enhanced FlowVisor}
\label{EnFV:sec}
Enhanced FlowVisor~\cite{Min2012} extends FlowVisor
addressing and tackles FlowVisor's simple bandwidth allocation and lack of
admission control. Enhanced FlowVisor is
implemented as an extension to the NOX SDN controller~\cite{gud2008nox}.
Enhanced FlowVisor uses VLAN PCP~\cite{sof2009sur} to achieve
flow-based bandwidth guarantees.
Before a virtual network request is accepted, the admission
control module checks whether enough link capacity is available. In
case the residual link bandwidth is not sufficient, a virtual network
request is rejected.

\subsubsection{Slices Isolator}
Slices Isolator~\cite{El-Azzab2011}
mainly focuses on concepts providing isolation between virtual slices
sharing an SDN switch.
Slice Isolator is implemented as a software
extension of the hypervisor layer, which is positioned between
the physical SDN network and the virtual SDN controllers, as illustrated in
Fig.~\ref{concepts:fig}(b).
The main goal of Slices Isolator is to adapt to the isolation demands of
the virtual network users.
Slices Isolator introduces an isolation model for NE resources with
eight isolation levels. Each isolation level is
a combination of activated or
deactivated isolation of the main NE resources, namely
interfaces, packet processing, and buffer memory.
The lowest isolation level does not isolate any of
these resources, i.e., there is no interface isolation, no packet
processing isolation, and no buffer memory isolation. The highest
level isolates interfaces, packet processing, and buffer memory.

If multiple slices (vSDNs) share a physical interface (port),
Slices Isolator provides interface isolation by mapping incoming
packets to the corresponding slice processing pipeline.
Packet processing isolation is implemented via allocating flow
tables to vSDNs. In order to improve processing, Slices Isolator
introduces the idea of sharing flow tables for common operations. For
example, two vSDNs operating only on Layer~3 can share a flow table
that provides ARP table information. Memory isolation targets the
isolation of shared buffer for network packets. If memory isolation is
required, a so-called traffic manager sets up
separated queues, which guarantee memory isolation.

\subsubsection{Double-FlowVisors}
The Double-FlowVisors approach~\cite{Yin2013} employs two instances of
FlowVisor. The first FlowVisor sits between the vSDN controllers and the
physical SDN network
(i.e., in the position of the Hypervisor in Fig.~\ref{concepts:fig}(b)).
The second FlowVisor is positioned between the vSDN controllers and the
applications.
This second FlowVisor gives applications a unified view of the
vSDN controller layer, i.e., the second FlowVisor virtualizes (abstracts)
the vSDN control.

\subsection{Hypervisors for Special Network Types}
\label{specenhyp:sec}
In this section, we survey the hypervisors that have to date been
investigated for special network types. We first cover wireless and
mobile networks, followed by optical networks, and then enterprise
networks.

\subsubsection{CellVisor} \label{CellVisor:sec}
CellSDN targets an architecture for cellular core networks that
  builds on SDN in order to simplify network control and
  management~\cite{li2012tow,Li}.
Network virtualization is an important aspect
  of the CellSDN architecture and is achieved through an integrated CellVisor
   hypervisor that is an extension of FlowVisor.
  In order to manage and control the
  network resources according to subscriber demands,
  CellVisor flexibly slices the wireless network resources, i.e.,
  CellVisor slices the base stations and radio resources.
Individual controllers manage and control the radio resources for
the various slices according to subscriber demands. The controllers
conduct admission control for the sliced resources and provide
mobility. CellVisor extends FlowVisor through the new feature of
slicing the base station
  resources.
Moreover, CellVisor adds a so called ''slicing of the semantic
  space''. The semantic space encompasses all subscribers whose packets
  belong to the same classification. An example classification could
  be traffic of all roaming subscribers or all subscribers from a
  specific mobile Internet provider. For differentiation, CellVisor uses
  MPLS tags or VLAN tags.

\subsubsection{RadioVisor}
\label{RadV:sec}
The conceptual structure and operating principles of a RadioVisor for
sharing radio access networks are presented in~\cite{Gudipati2014,gud2014rad}.
RadioVisor considers a three dimensional (3D) resource grid
consisting of space (represented through spatially distributed
radio elements), time slots, and frequency slots.
The radio resources in the 3D resource grid are sliced by RadioVisor to
enable sharing by different controllers. The controllers in turn provide
wireless services to applications.
RadioVisor periodically (for each time window with a prescribed duration)
slices the 3D resource grid to assign
radio resources to the controllers.
The resource allocation is based on the current (or predicted) traffic load
of the controllers and their service level agreement with RadioVisor.
Each controller is then allowed to control its allocated radio resources
from the 3D resource grid for the duration of the current time window.
A key consideration for RadioVisor is that the radio resources
allocated to distinct controllers should have isolated wireless
communication properties. Each controller should be able to
independently utilize its allocated
radio resources without coordinating with other controllers.
Therefore, RadioVisor can allocate the same time and frequency slot to
multiple spatially distributed radio elements only if the radio elements
are so far apart that they do not interfere with each other when
simultaneously transmitting on the same frequency.

\subsubsection{MobileVisor}
The application of the FlowVisor approach for mobile networks is outlined
through the overview of a MobileVisor architecture in~\cite{Nguyen2014}.
MobileVisor integrates the FlowVisor functionality into the
architectural structure of a virtual mobile packet network that can potentially
consist of multiple underlying physical mobile networks, e.g.,
a 3G and a 4G network.

\subsubsection{Optical FlowVisor}
For the context of an SDN-based optical network~\cite{Chan2013},
the architecture and initial experimental results for an Optical FlowVisor
have been presented in~\cite{Azodolmolky2012}.
Optical FlowVisor employs the FlowVisor principles to create
virtual optical networks (VONs) from an underlying optical circuit-switched
network.
Analogous to the consideration of the wireless
communication properties in RadioVisor (see Section~\ref{RadV:sec}),
Optical FlowVisor needs to consider the physical layer
impairments of the optical communication channels
(e.g., wavelength channels in a WDM network~\cite{ish1984rev}).
The VONs should be constructed such that each controller can
utilize its VON without experiencing significant physical layer
impairments due to the transmissions on other VONs.

\subsubsection{EnterpriseVisor}
For an enterprise network with a specific configuration an
EnterpriseVisor is proposed in~\cite{Chen2014a} to complement the
operation of the conventional FlowVisor.
The EnterpriseVisor operates software modules in the hypervisor layer
to monitor and analyze the network deployment.
Based on the network configuration stored in a database in the
EnterpriseVisor, the goal is to monitor the FlowVisor operation
and to assist FlowVisor with resource allocation.

\subsection{Policy-based Hypervisors}
\label{polcenhyp:sec}
In this section, we survey hypervisors that focus on
supporting heterogeneous SDN controllers (and their corresponding
network applications) while providing the advantages of
virtualization, e.g., abstraction and simplicity of management.
The policy-based hypervisors compose OF rules for operating
SDN switches from the inputs of multiple distinct network applications
(e.g., a firewall application and a load balancing application)
and corresponding distinct SDN controllers.

\subsubsection{Compositional Hypervisor}
The main goal of the Compositional Hypervisor~\cite{Rexford2014} is to
provide a platform that allows SDN network operators to choose network
applications developed for different SDN controllers.
Thereby, the Compositional
Hypervisors gives network operators the flexibility to choose from a
wide range of SDN applications, i.e., operators are not limited to the
specifics of only one particular SDN controller
(and the network applications supported by the controller).
The Compositional Hypervisor enables SDN controllers to ``logically''
cooperate on the same traffic.
For conceptual illustration, consider Fig.~\ref{concepts:fig}(b)
and suppose that $\mbox{App}_{11}$ is a firewall application written
for the vSDN 1 Controller, which is the Python controller Ryu~\cite{ryu}.
Further suppose that $\mbox{App}_{21}$ is a load-balancing application written
for the vSDN 2 Controller, which is the C++ controller NOX~\cite{gud2008nox}.
The Compositional Hypervisor allows these two distinct network applications
through their respective vSDN controllers to logically operate on the
same traffic.

Instead of strictly isolating the
traffic, the Compositional Hypervisor forms a ``composed policy'' from the
individual policies of the multiple vSDN controllers.
More specifically, a policy represents a prioritized list of
OF rules for operating each SDN switch.
According to the network applications, the vSDN controllers give
their individual policies to the Compositional Hypervisor.
The Compositional Hypervisor, in turn, prepares a composed policy,
i.e., a composed prioritized list of OF rules for the SDN switches.
The individual policies are composed according to a composition configuration.
For instance, in our firewall and load balancing example, a reasonable
composition configuration may specify that the firewall rules have
to be processed first and that the load-balancing rules are not
allowed to overwrite the firewall rules.

The Compositional Hypervisor evaluation in~\cite{Rexford2014} focuses
on the overhead for the policy composition.
The Compositional Hypervisor needs to update the composed policy when a
vSDN controller wants to add, update, or delete an OF rule.
The overhead is measured in terms of the computation overhead
and the rule-update overhead.
The computation overhead is the time for calculating the new flow table,
which has to be installed on the switch.
The rule-update overhead is the amount of messages that
have to be send to convert the old flow table state of a switch to
the newly calculated flow table.

\subsubsection{CoVisor}
CoVisor~\cite{Jin2015} builds on the concept of the Compositional
Hypervisor.
Similar to the Compositional Hypervisor, CoVisor
focuses on the cooperation of heterogeneous SDN controllers,
i.e., SDN controllers written in different programming languages,
on the same network traffic.
While the Compositional Hypervisor study~\cite{Rexford2014}
focused on algorithms for improving the policy composition process, the
CoVisor study~\cite{Jin2015} focuses on improving the performance of the
physical SDN network, e.g., how to compose OF rules to save flow table space
or how to abstract the physical SDN topology to improve the performance of
the vSDN controllers.
Again, the policies (prioritized OF rules) from multiple network applications
running on different vSDN controllers are composed to a single
composed policy, which corresponds to a flow table setting. The single
flow table still has to be correct, i.e., to work as if
no policy composition had occurred.
Abstraction mechanisms are developed in order to provide a vSDN controller
with only the ``necessary'' topology information. For instance, a firewall
application may not need a detailed view of the underlying topology in
order to decide whether packets should be dropped or forwarded.
Furthermore, CoVisor provides mechanisms to protect
the physical SDN network against malicious or buggy vSDN controllers.

The CoVisor performance evaluation in~\cite{Jin2015} compares the CoVisor
policy update algorithm with the Compositional Hypervisor algorithm
The results show improvements on the order of two to three
orders of magnitude due to the more sophisticated flow table updates
and the additional virtualization capabilities (topology abstraction).

\section{Distributed Hypervisors}
\label{distr:sec}
\subsection{Execution on General Computing Platform}
\label{gcpdishyp:sec}
\subsubsection{FlowN} \label{FlowN:sec}
FlowN~\cite{Drutskoy2013b} is a distributed hypervisor for
virtualizing SDN networks. However, tenants cannot employ their own
vSDN controller. Instead, FlowN provides a container-based application
virtualization. The containers host the tenant controllers,
which are an extension of the NOX controller. Thus, FlowN
users are limited to the capabilities of the NOX controller.

Instead of only slicing the physical network, FlowN completely
abstracts the physical network and provides virtual network topologies
to the tenants. An advantage of this abstraction is, for example, that
virtual nodes can be transparently migrated on the physical network.
Tenants are not aware of these resource management
actions as they see only their virtual topologies. Furthermore, FlowN
presents only virtual address spaces to the tenants. Thus, FlowN always has to
map between virtual and physical address spaces. For this
purpose, FlowN uses an additional data base component for
providing a consistent mapping. For scalability, FlowN relies on a
master-slave database principle. The state of the master database is
replicated among multiple slave databases. Using this database
concept, FlowN can be distributed among multiple physical
servers. Each physical server can be equipped with a database and a
controller, which is hosting a particular number of containers, i.e.,
controllers.

To differentiate between vSDNs, edge switches encapsulate and
decapsulate the network packets with VLAN tagging.
Furthermore, each tenant gets only a pre-reserved amount
of available flowspace on each switch.  In order to provide resource
isolation between the tenant controllers, FlowN assigns one processing
thread per container.

The evaluation in~\cite{Drutskoy2013b} compares the FlowN architecture
with two databases with FlowVisor in terms of the hypervisor latency overhead.
The number of virtual networks is increased from $0$ to $100$.
While the latency overhead of FlowVisor
increases steadily, FlowN always shows a constant latency overhead. However,
the latency overhead of FlowVisor is lower for $0$ to $80$
virtual networks than for FlowN.

\subsubsection{Network Hypervisor}
The Network Hypervisor~\cite{hua2013net} addresses the challenge of
``stitching'' together a vSDN slice from different underlying physical SDN
infrastructures (networks).
The motivation for the Network Hypervisor is the complexity
of SDN virtualization arising from
the current heterogeneity of SDN infrastructures.
Current SDN infrastructures provide different levels of abstraction
and a variety of SDN APIs.
The proposed Network Hypervisor mainly contributes to
the abstraction of multiple SDN infrastructures as a virtual slice.
Similar to FlowN, the Network Hypervisor acts as a controller to the
network applications of the vSDN tenants.
Network applications can use a
higher-level API to interact with the Network Hypervisor, while the
Network Hypervisor interacts with the different SDN
infrastructures and complies with their respective API attributes.
This Network Hypervisor design provides
vSDN network applications with a transparent operation of multi-domain SDN
infrastructures.

A prototype Network Hypervisor was implemented
on top of GENI testbed and supports the GENI API~\cite{ber2014gen}.
The demonstration in~\cite{hua2013net} mapped a virtual SDN slice across both
a Kentucky Aggregate and a Utah Aggregate on the GENI testbed.
The Network Hypervisor fetches resource and topology information from both
aggregates via the discover API call.

\subsubsection{AutoSlice}
AutoSlice~\cite{Bozakov2012, Bozakov2014} strives to improve the
scalability of a logically centralized hypervisor by
distributing the hypervisor workload.
AutoSlice targets software deployment on general-purpose computing platforms.
The AutoSlice concept segments the physical
infrastructure into non-overlapping SDN domains.
The AutoSlice hypervisor is partitioned into a single management module
and multiple controller proxies, one proxy for each SDN physical domain.
The management module assigns the virtual resources to the proxies.
Each proxy stores the resource assignment and translates the
messages exchanged between the vSDN controllers and the physical
SDN infrastructure in its domain.
Similar to FlowVisor, AutoSlice is positioned
between the physical SDN network and the vSDN controllers,
see Fig.~\ref{concepts:fig}(b).
AutoSlice abstracts arbitrary SDN topologies, processes and rewrites control
messages, and enables SDN node and link migration.
In case of migrations due to
substrate link failures or vSDN topology changes,
AutoSlice migrates the flow tables and the
affected network traffic between SDN switches in the correct update sequence.

Regarding isolation, a partial control plane offloading could be
offered by distributing the hypervisor over multiple proxies.  For
scalability, AutoSlice deploys auxiliary software
datapaths (ASDs)~\cite{Yu2010}. Each ASD is equipped with a software switch,
e.g., OpenVSwitch (OVS)~\cite{pfa2009ext}, running on a commodity server.
Commodity servers have plentiful memory and thus can cache the full
copies of the OF rules.  Furthermore, to improve scalability on the
virtual data plane, AutoSlice differentiates between mice and elephant
network flows~\cite{guo2001war,lan2006mea,sou2004flo}.
Mice flows are cached in the corresponding ASD
switches, while elephant flows are stored in the dedicated switches.
AutoSlice uses a so-called virtual flow table identifier (VTID)
to differentiate flow
table entries of vSDNs on the SDN switches.  For realization of the
VTIDs, AutoSlice assumes the use of MPLS~\cite{xia2000tra}
or VLAN~\cite{sof2009sur} techniques.  Using
this technique, each vSDN receives the full flowspace.  Within each
SDN domain, the control plane isolation problem still persists.
The AutoSlice studies~\cite{Bozakov2012, Bozakov2014} do not provide any
performance evaluation, nor demonstrate a prototype implementation.

\subsubsection{NVP}
The Network Virtualization Platform (NVP)~\cite{Koponen2014} targets the
abstraction of data center network resources to be managed by cloud tenants.
In today's multi-tenant data centers, computation and storage have long
been successfully abstracted by computing hypervisors,
e.g., VMware~\cite{ros1999vmw,wal2002mem} or KVM~\cite{kiv2007kvm}.
However, tenants have typically not been given the ability to manage
the cloud's networking resources.

Similar to FlowN, NVP does not allow tenants to run their own controllers.
NVP acts as a controller that provides the tenants' applications with an
API to manage their virtual slice in the data center. NVP wraps the ONIX
controller platform~\cite{kop2010oni}, and thus inherits the distributed
controller architecture of ONIX. NVP can run a distributed controller cluster
within a data center to scale to the operational load and requirements of
the tenants.

NVP focuses on the virtualization of the SDN software switches,
e.g., Open vSwitches (OVSs)~\cite{pfa2015des}, that steer the traffic
to virtual machines (VMs),  i.e., NVP focuses on the software switches
residing on the host servers.
The network infrastructure in the data center between servers is not
controlled by NVP, i.e., not virtualized.
The data center physical network is assumed to provide a uniform balanced
capacity, as is common in current data centers.
In order to virtualize (abstract)
the physical network, NVP creates logical datapaths,
i.e., overlay tunnels, between the source and destination OVSs.
Any packet entering the host OVS, either from
a VM or from the overlay tunnel is sent through a logical pipeline
corresponding to the logical datapath to which the packet belongs.

The logical paths and pipelines abstract the data plane for
the tenants, whereby a logical path and a logical pipeline are assigned for
each virtual slice.
NVP abstracts also the control plane, whereby a tenant can set the routing
configuration and protocols to be used in its virtual slice.
Regarding isolation, NVP assigns flow tables from the OVSs
to each logical pipeline with a unique identifier.
This enforces isolation from other logical datapaths and places the
lookup entry at the proper  logical pipeline.

The NVP evaluation in~\cite{Koponen2014} focuses on the concept of
logical datapaths, i.e., tunnels for data plane virtualization.
The throughput of two encapsulation methods for tunneling,
namely Generic Routing Encapsulation (GRE)~\cite{han2000gen} and
State-less Transport Tunneling (STT)~\cite{gro2014stt}, is compared to
a non-tunneled (non-encapsulated) benchmark scenario.
The results indicate that STT achieves approximately
the same throughput as the non-tunneled benchmark, whereas GRE
encapsulation reduces the throughput to less than a third
(as GRE does not employ hardware offloading).

\subsection{Computing Platform $+$ General-Purpose NE based}
\label{gcpgpne:sec}
\subsubsection{OpenVirteX}
OpenVirteX~\cite{Al-Shabibi2014, Berde2014, Al-shabibi2014b} provides
two main contributions: address virtualization and topology virtualization.
OpenVirteX builds on the design of
FlowVisor, and operates (functions) as an intermediate layer between vSDNs
and controllers.

OpenVirteX tackles the so-called flowspace problem:
The use of OF header fields to distinguish vSDNs prevents
hypervisors from offering the entire OF header fields space to the vSDNs.
OpenVirteX provides each tenant with the full header fields space.
In order to achieve this, OpenVirteX places edge switches at the borders of
the physical SDN network. The edge switches re-write
the virtually assigned IP and MAC addresses, which are used by the hosts
of each vSDN (tenant), into
disjoint addresses to be used within the physical SDN network.
The hypervisor ensures that the correct address
mapping is stored at the edge switches. With this mapping approach,
the entire flowspace can be provided to each vSDN.

To provide topology abstraction, OpenVirteX does not operate completely
transparently (compared to the transparent FlowVisor operation).
Instead, as OpenVirteX knows the
exact mapping of virtual to physical networks, OpenVirteX answers
Link Layer Discovery Protocol (LLDP)~\cite{802.1ab} controller messages
(instead of the physical SDN switches).
No isolation concepts, neither for the data plane,
 nor for the control plane, have been presented.
In addition to topology customization, OpenVirteX provides an
advanced resilience feature based on its topology virtualization
mechanisms. A virtual link can be mapped to multiple physical links;
vice versa, a virtual SDN switch can be realized by multiple
physical SDN counterparts.

The OpenVirteX evaluation in~\cite{Al-Shabibi2014, Berde2014, Al-shabibi2014b}
compares the control plane latency overheads of OpenVirteX, FlowVisor,
FlowN, and a reference case without virtualization.
For benchmarking, Cbench~\cite{ToGGC12} is used and five switch instances are
created. Each switch serves a specific number of hosts, serving as
one virtual network per switch. Compared to the other hypervisors,
OpenVirteX achieves better performance and adds a latency of only
0.2~ms compared to the reference case. Besides the
latency, also the instantiation time of virtual networks was benchmarked.

\subsubsection{OpenFlow-based Virtualization Framework for the Cloud
(OF NV Cloud)}
OF NV Cloud~\cite{Matias2011} addresses virtualization of
data centers, i.e., virtualization inside a data center and
virtualization of the data center interconnections.
Furthermore, OF NV Cloud addresses the virtualization of multiple physical
infrastructures.
OF NV Cloud uses a MAC addressing scheme for address virtualization.
In order to have unique addresses in the data centers, the MAC addresses
are globally administered and assigned.
In particular, the MAC address is divided into a 14~bit vio\_id
(virtual infrastructure operator) field,
a 16~bit vnode\_id field, and a 16~bit vhost\_id field. In order
to interconnect data centers, parts of the vio\_id field are
reserved. The vio\_id is used to identify vSDNs (tenants).
The resource manager, i.e., the entity responsible for
assigning and managing the unique MAC addresses as well as for access to
the infrastructure, is similarly designed as FlowVisor.
However, the OF NV Cloud virtualization cannot be implemented for
OF~1.1 as it does not provide rules based on MAC prefixes. In terms of
data plane resources, flow tables of switches are split among
vSDNs. No evaluation of the OF NV Cloud architecture has been reported.

\subsubsection{AutoVFlow}
AutoVFlow~\cite{Yamanaka2014,Yamanaka} is a distributed hypervisor for
SDN virtualization.
In a wide-area network, the infrastructure is divided into
non-overlapping domains. One possible distributed hypervisor deployment
concept has a proxy responsible for each domain and a central hypervisor
administrator that configures the proxies.
The proxies would only act as containers, similar to
the containers in FlowN, see Section~\ref{FlowN:sec}, to enforce the
administrator's configuration, i.e., abstraction and control plane
mapping, in their own domain.
The motivation for AutoVFlow is that such a distributed
structure would highly load the central administrator. Therefore, AutoVFlow
removes the central administrator and delegates the configuration role
to distributed administrators, one for each domain. From an
architectural point-of-view, AutoVFlow adopts a flat distributed
hypervisor architecture without hierarchy.

Since a virtual slice can span multiple domains, the distributed
hypervisor administrators need to exchange the virtualization
configuration and policies among each other in order to maintain the
slice state. AutoVFlow uses virtual identifiers, e.g., virtual MAC
addresses, to identify data plane packet flows from different
slices. These virtual identifiers can be different from one domain to
the other. In case a packet flow from one slice is entering another
domain, the hypervisor administrator of the new domain
identifies the virtual slice based on
the identifier of the flow in the previous domain.
Next, the administrator of the new domain
replaces the identifier from the previous domain by the
identifier assigned in the new domain.

The control plane latency overhead induced by adding AutoVFlow,
between the tenant controller and the SDN network, has been
evaluated in~\cite{Yamanaka2014,Yamanaka}.
The evaluation setup included two domains with two AutoVFlow
administrators. The latency was measured for OF
PCKT\_IN, PCKT\_OUT, and FLOW\_MOD messages. The highest impact was
observed for FLOW\_MOD messages, which experienced a latency overhead
of 5.85~ms due to AutoVFlow.

\subsection{Computing Platform $+$ Special-Purpose NE-based}
\label{gcpspne:sec}
\subsubsection{Carrier-grade}
A distributed SDN virtualization architecture referred to
as Carrier-grade has been introduced in~\cite{Devlic2012a,Skoldstrom2013}.
Carrier-grade extends the physical SDN hardware in order to realize
vSDN networks.
On the data plane, vSDNs are created and differentiated via labels,
e.g., MPLS labels, and the partitioning of the flow tables. In order to
demultiplex encapsulated data plane packets and to determine the
corresponding vSDN flow tables, a virtualization controller controls a
virtualization table. Each arriving network packet is first matched
against rules in the virtualization table. Based, on the flow table entries for
the labels, packets are forwarded to the corresponding vSDN flow
tables. For connecting vSDN switches to vSDN controllers,
Carrier-grade places translation units in every physical SDN
switch. The virtualization controller provides the set of policies and
rules, which include the assigned label, flow table, and port for each
vSDN slice, to the translation units.  Accordingly, each vSDN
controller has only access to its assigned flow tables. Based on the
sharing of physical ports, Carrier-grade uses different technologies,
e.g., VLAN and per port queuing, to improve data plane performance and
service guarantees.

The distributed Carrier-grade architecture aims at minimizing the
overhead of a logically centralized hypervisor by providing direct
access from the vSDN controllers to the physical infrastructure.
However, Carrier-grade adds processing complexity to the
physical network infrastructure.
Thus, the Carrier-grade evaluation~\cite{Devlic2012a,Skoldstrom2013} examines
the impact of the additional encapsulation in the data plane on the
latency and throughput. As a baseline setup with only one
virtual network running on the data plane is considered, conclusions about the
performance in case of interfering vSDNs and overload scenarios cannot
be given. Carrier-grade adds a relative latency of 11~\%.

The evaluation in~\cite{Devlic2012a,Skoldstrom2013} notes that the
CPU load due to the added translation unit should be negligible.
However, a deeper analysis has not been provided.
The current Carrier-grade design does not specify how the available
switch resources (e.g., CPU and memory) are used by the translation
unit and are shared among multiple tenants. Furthermore, the
scalability aspects have to be investigated in detail in future research.

\subsubsection{Datapath Centric}
Datapath Centric~\cite{DoriguzziCorin2014} is a hypervisor designed to
address the single point of failure in the
FlowVisor design~\cite{Sherwood2009} and to improve
the performance of the virtualization layer by implementing
virtualization functions as switch extensions.
Datapath Centric is based on the eXtensible Datapath Daemon (xDPd) project,
which is an open-source datapath project~\cite{xdpd}. xDPd in its
used versions supports OF~1.0 and OF~1.2. Relying on xDPd, Datapath
Centric simultaneously supports different OF switch versions, i.e., it
can virtualize physical SDN networks that are composed of switches
supporting OF versions from 1.0 to 1.2.

The Datapath Centric architecture consists of the Virtualization
Agents (VAs, one in each switch)
and a single Virtualization Agent Orchestrator (VAO).
The VAO is responsible for the slice configuration and monitoring,
i.e., adding slices, removing slices, or extending flowspaces of
existing slices. The VAs implement the distributed slicing
functions, e.g., the translation function. They directly communicate
with the vSDN controllers. Additionally, the VAs abstract the
available switch resources and present them to the VAO.  Due to the
separation of VAs and VAO, Datapath Centric avoids the single point of
failure in the FlowVisor architecture. Even in
case the VAO fails, the VAs can continue operation. Datapath Centric
relies on the flowspace concept as introduced by FlowVisor. Bandwidth
isolation on the data plane is not available but its addition is in the
planning based on the QoS capabilities of xDPd.

The latency overhead added by the VA agent has been
evaluated for three cases~\cite{DoriguzziCorin2014}:
a case only running xDPd without the VA (reference),
a case where the VA component is added,
and a final case where FV is used. The VA case adds an additional overhead
of 18~\% compared to the reference case. The gap between the VA
and FV is on average 0.429~ms. In a second
evaluation, the scalability of Datapath Centric was evaluated for
$500$ to $10000$ rules. It is reported that the additional
latency is constant from $100$ to $500$ rules, and then scales
linearly from $500$ to $10000$ rules, with a maximum of 3~ms. A
failure scenario was not evaluated. A demonstration of Datapath
Centric was shown in~\cite{dep2014dem}.

\subsubsection{DFVisor}
Distributed FlowVisor (DFVisor)~\cite{Liao2014,Liao2015}
is a hypervisor designed to address
the scalability issue of FlowVisor as a centralized SDN virtualization
hypervisor.  DFVisor realizes the virtualization layer on the SDN
physical network itself.  This is done by extending the SDN switches
with hypervisor capabilities, resulting in so-called
``enhanced OpenFlow switches''.  The SDN switches are extended by a
local vSDN slicer and tunneling module. The slicer module implements
the virtual SDN abstraction and maintains the slice configuration on
the switch. DFVisor uses GRE tunneling~\cite{han2000gen}
to slice the data plane and
encapsulate the data flows of each slice.
The adoption of GRE tunneling is motivated by the
use of the GRE header stack to provide slice QoS.

DFVisor also includes a distributed synchronized two-level
database system that consists of local databases on the switches and a
global database. The global database maintains the virtual slices
state, e.g., flow statistics, by synchronizing with the local
databases residing on the switches. This way the global database can
act as an interface to the vSDN controllers. Thereby, vSDN controllers do
not need to access individual local databases, which are distributed on
the switches. This can facilitate the network operation and improve
the virtualization scalability.

\subsubsection{OpenSlice}
OpenSlice~\cite{liu2013os,liu2011exp} is a hypervisor design for
elastic optical networks (EONs)~\cite{ger2012ela,jin2009spe}.
EONs adaptively allocate the optical communications spectrum to
 end-to-end optical paths so as to achieve different transmission bit rates
and to compensate for physical layer optical impairments.
The end-to-end optical path are routed through the EON by
distributed bandwidth-variable wavelength cross-connects
(BV-WXCs)~\cite{jin2009spe}.
The OpenSlice architecture interfaces the optical layer (EON)
with OpenFlow-enabled IP packet routers
through multi-flow optical transponders (MOTPs)~\cite{tak2011exp}.
A MOTP identifies packets and maps them to flows.

OpenSlice extends the OpenFlow protocol messages to carry
the EON adaption parameters, such as optical central frequency,
slot width, and modulation format.
OpenSlice extends the conventional MOTPs and BV-WXCs to
communicate the EON parameters through the extended OpenFlow protocol.
OpenSlice furthermore extends the conventional NOX controller to
perform routing and optical spectrum assignment. The extended
OpenSlice NOX controller assigns optical frequencies, slots, and
modulation formats according to the traffic flow requirements.

The evaluation in~\cite{liu2013os} compares the path provisioning latency
of OpenSlice with Generalized Multiple Protocol Label
Switching (GMPLS)~\cite{mun2009cha}.
The reported results indicate that the
centralized SDN-based control in OpenSlice has nearly constant
latency as the number of path hops increases, whereas the
hop-by-hop decentralized GMPLS control leads to increasing latencies with
increasing hop count.
Extensions of the OpenSlice concepts to support a wider range of
optical transport technologies and more flexible virtualization
have been examined
in~\cite{Chan2013,lop2015dem,mun2015sdn,mun2015int,nej2015sdn,szy2014dem,szy2015tow,vil2015net,vil2015mul}.

\subsubsection{Advanced Capabilities}
 The Advanced Capabilities OF virtualization
framework~\cite{Sonkoly2012b} is designed to provide SDN virtualization
with higher levels of flexibility than FlowVisor~\cite{Sherwood2009}.
First, FlowVisor requires all
SDN switches to operate with the same version of the OF protocol.
The Advanced Capabilities framework can manage vSDNs that
consist of SDN switches running different OF versions.
The introduced framework also accommodates flexible forwarding mechanisms
and flow matching in the individual switches.
Second, the Advanced Capabilities framework includes a network management
framework that can control the different distributed vSDN controllers.
The network management framework can centrally monitor and maintain
the different vSDN controllers.
Third, the Advanced Capabilities framework can configure queues in the
SDN switches so as to employ QoS packet scheduling mechanisms.
The Advanced Capabilities framework implements the outlined
features through special management modules that are
inserted in the distributed SDN switches.
The specialized management modules are based on the
network configuration protocol~\cite{enn2011net} and use the
YANG data modeling language~\cite{bjo2010yang}.
The study~\cite{Sonkoly2012b} reports on a proof-of-concept prototype
of the Advanced Capabilities framework, detailed quantitative performance
measurements are not provided.

\subsection{Computing Platform $+$ Special- and General-Purpose NE-based}
\label{gcpgspne:sec}
\subsubsection{HyperFlex}
HyperFlex~\cite{ble2015hyp} introduces the idea of realizing the
hypervisor layer via multiple different virtualization
functions. Furthermore, HyperFlex explicitly addresses the control plane
virtualization of SDN networks.
It can operate in a centralized or distributed manner.
It can
realize the virtualization according to the available capacities of
the physical network and the commodity server platforms. Moreover, HyperFlex
operates and interconnects the functions needed for virtualization,
which as a whole realize the hypervisor layer.
In detail, the HyperFlex concept
allows to realize functions in software, i.e., they can be placed and
run on commodity servers, or in hardware, i.e., they can be realized via
the available capabilities of the physical networking (NE) hardware.
The HyperFlex design thus increases the flexibility and
scalability of existing hypervisors. Based on the current vSDN
demands, hypervisor functions can be adaptively scaled.
This dynamic adaptation provides a fine resource management granularity.

The HyperFlex concept has initially been realized and demonstrated
for virtualizing the control plane of SDN networks.
The software isolation function operates on the application layer, 
dropping OF messages that exceed a prescribed vSDN message rate.
The network isolation function operates on layers 2--4, 
policing (limiting) the vSDN control channel rates on NEs.
It was shown that control plane virtualization functions
either realized in software or in hardware can isolate vSDNs.
In case the hypervisor layer is over-utilized, e.g., its CPU
is completely utilized, the performance of several or all vSDN slices
can be degraded, even if only one vSDN is responsible for the over-utilization.
In order to avoid hypervisor over-utilization due to a
tenant, the hypervisor CPU has to be sliced as well. This means
that specific amounts of the available CPU resources are assigned to the
tenants. In order to quantify the relationship between CPU and control
plane traffic per tenant, i.e., the amount of CPU resources needed to
process the tenants' control traffic, the hypervisor has to be
benchmarked first. The benchmarking measures, for example, the average
CPU resource amount needed for the average control plane traffic, i.e., OF
control packets. 
The hypervisor isolation functions are then configured according to
the benchmark measurements, i.e., they are set to support a
guaranteed OF traffic rate.

In order to achieve control plane isolation, software and hardware
isolation solutions exhibit trade-offs, e.g., between OF control
message latency and OF control message loss. 
More specifically, a hardware solution 
polices (limits) the OF control traffic on the network layer, 
for instance, through shapers (egress) or policers (ingress) of 
general-purpose or special-purpose NEs. 
However, the shapers do not specifically 
drop OF messages, which are application layer messages. 
This is because current OF switches cannot match and drop application layer 
messages (dropping specifically OF messages would require a proxy, 
as control channels may be encrypted). 
Thus, hardware solutions simply drop network packets based on matches up to 
layer 4 (transport layer).
In case of using TCP as the control channel transmission protocol, 
the retransmissions of dropped packets lead to an
increasing backlog of buffered packets in the sender and NEs; 
thus, increasing control
message latency. On the other hand, the software solution drops OF
messages arriving from the network stack in order to reduce
the workload of subsequent hypervisor functions. This, however, means
a loss of OF packets, which a tenant controller would have to compensate for.
A demonstration of the HyperFlex architecture with its
hardware and software isolation functions was provided in~\cite{bas2015hyp}.

\section{Summary Comparison of Surveyed Hypervisors}
\label{sec:summary}
\begin{center}
\begin{table*}[t]
    \centering
    \caption{Summary comparison of abstraction of physical network attributes
       by existing hypervisors (the check mark indicates that a network
   attribute is virtualized; whereas ``--'' indicates that a
   network attribute is not virtualized).}
\label{tab:abstraction_comparison}
    \begin{tabular}{|l|l|l|p{3cm}|l|l|l|p{2cm}| }
    \hline
\multicolumn{3}{|l|}{Classification} & Hypervisor & Topology & Physical Node Resources
                              & Physical Link Resources \\ \hline \hline
\multirow{15}{*}{Centr.} &
\multirow{15}{*}{\begin{minipage}{0.3in} Gen. Comp. Platf. \end{minipage}}&
 \multirow{7}{*}{\begin{minipage}{0.4in} Gen., Secs.~\ref{FlowVisor:sec}, \ref{cenotherhyp:sec} \end{minipage}} &
FlowVisor~\cite{Sherwood2009} & -- & -- & -- \\ \cline{4-7}
& & & ADVisor~\cite{Salvadori2011} & -- & \checkmark & -- \\ \cline{4-7}
& & & VeRTIGO~\cite{Corin2012} & \checkmark & \checkmark & -- \\ \cline{4-7}
& & & Enhanced FlowVisor~\cite{Min2012} & -- & -- & \checkmark \\ \cline{4-7}
& & & Slices Isolator~\cite{El-Azzab2011} & -- & \checkmark & \checkmark \\ \cline{4-7}
& & & Double FlowVisor~\cite{Yin2013} & -- & -- & -- \\ \cline{3-7}

& & \multirow{5}{*}{\begin{minipage}{0.3in} Spec. Netw. Types, Sec.~\ref{specenhyp:sec} \end{minipage}} & 
CellVisor~\cite{li2012tow,Li} & -- & \checkmark (base station) & \checkmark (radio resources) \\ \cline{4-7}
& & & RadioVisor~\cite{Gudipati2014} & -- & -- & \checkmark (radio link) \\ \cline{4-7}
& & & MobileVisor~\cite{Nguyen2014} & -- & -- & -- \\ \cline{4-7}
& & & Optical FV~\cite{Azodolmolky2012} & -- & \checkmark & \checkmark (optical link) \\ \cline{4-7}
& & & Enterprise Visor\cite{Chen2014a} & -- & -- & -- \\ \cline{3-7}
& & \multirow{2}{*}{\begin{minipage}{0.5in} Policy-b., Sec.~\ref{polcenhyp:sec} \end{minipage}} & \multicolumn{1}{l|}{Compositional Hypervisor~\cite{Rexford2014}}  & -- & -- & -- \\ \cline{4-7}
& & & CoVisor~\cite{Jin2015} & \checkmark & \checkmark  & -- \\ \hline \hline

\multirow{20}{*}{Distr.} &
\multicolumn{2}{l|}
{\multirow{5}{*}
    {\begin{minipage}{0.8in}Gen. Comp. Platf., Sec.~\ref{gcpdishyp:sec}
            \end{minipage}}} &
FlowN~\cite{Drutskoy2013b} & \checkmark & -- & -- \\ \cline{4-7}
& \multicolumn{2}{l|}{} & Network Hypervisor~\cite{hua2013net}  & \checkmark & -- & -- \\ \cline{4-7}
& \multicolumn{2}{l|}{} & AutoSlice~\cite{Bozakov2012,Bozakov2013} & \checkmark & -- & -- \\ \cline{4-7}
& \multicolumn{2}{l|}{} & Network Virtualization Platform (NVP)~\cite{Koponen2014} & -- & \checkmark & \checkmark \\ \cline{2-7}

& \multicolumn{2}{l|}{\multirow{3}{*}{\begin{minipage}{0.8in}Gen. Comp. Platf. + Gen.-purp. NE, Sec.~\ref{gcpgpne:sec}\end{minipage}}} & OpenVirteX~\cite{Al-Shabibi2014,Berde2014, Al-shabibi2014b} & \checkmark & -- & -- \\ \cline{4-7}
& \multicolumn{2}{l|}{} & OF NV Cloud~\cite{Matias2011} & -- & -- & -- \\ \cline{4-7}
& \multicolumn{2}{l|}{} & AutoVFlow~\cite{Yamanaka2014} & -- & -- & -- \\ \cline{2-7}

& \multicolumn{2}{l|}{\multirow{6}{*}{\begin{minipage}{0.8in}Gen. Comp. Platf. + Spec.-purp. NE, Sec.~\ref{gcpspne:sec}\end{minipage}}} & Carrier-grade~\cite{Devlic2012a} & -- & \checkmark & \checkmark \\  \cline{4-7}
& \multicolumn{2}{l|}{} & Datapath Centric~\cite{DoriguzziCorin2014} & -- & -- & -- \\ \cline{4-7}
& \multicolumn{2}{l|}{} &DFVisor~\cite{Liao2014,Liao2015} & -- & \checkmark & \checkmark \\ \cline{4-7}
& \multicolumn{2}{l|}{} & OpenSlice~\cite{liu2013os} & -- & -- & --  \\ \cline{4-7}
& \multicolumn{2}{l|}{} & Advanced Capabilities~\cite{Sonkoly2012b} & -- & -- & \checkmark \\ \cline{2-7}
& \multicolumn{2}{l|}
{\begin{minipage}{0.8in}
 Gen. Comp. Platf. + Gen.-purp. NE + Spec.-purp. NE, Sec.~\ref{gcpgspne:sec}
\end{minipage}} &
Hyperflex~\cite{ble2015hyp} & -- &  -- & -- \\  \hline   \hline
    \end{tabular}
\end{table*}
\end{center}

In this section, we present a comparison of the surveyed SDN
hypervisors in terms of the network abstraction attributes
(see Section~\ref{sec:abstraction}) and the isolation attributes
(see Section~\ref{sec:isolation}). The comparison gives a summary of
the differences between existing SDN hypervisors.
The summary comparison helps to place the
various hypervisors in context and to observe the focus areas and strengths,
but also the limitations of each proposed hypervisor.
We note that if an abstraction
or isolation attribute is not examined in a given hypervisor study, we
consider the attribute as not provided in the comparison.

\subsection{Physical Network Attribute Abstraction}
\label{sumabs:sec}
As summarized in Table~\ref{tab:abstraction_comparison},
SDN hypervisors can be differentiated according to the physical
network attributes that they can abstract and provide to their tenant
vSDN controllers in abstracted form.

\subsubsection{Topology Abstraction}
In a non-virtualized SDN network, the topology view assists the SDN
controller in determining the flow
connections and the network configuration. Similarly, a vSDN
controller requires a topology view of its vSDN to operate its
virtual SDN slice.
A requested virtual link in a vSDN slice could be
mapped onto several physical links.
Also, a requested virtual node could map to a combination of multiple
physical nodes. This abstraction of the
physical topology for the vSDN controllers, i.e., providing an
end-to-end mapping (embedding) of virtual nodes and links without exposing the
underlying physical mapping, can bring several advantages.
First, the operation of a vSDN slice is simplified, i.e, only
end-to-end virtual resources need to be controlled and operated.
Second, the actual physical topology is not
exposed to the vSDN tenants, which may alleviate security concerns
as well as concerns about revealing operational practices.
Only two existing centralized hypervisors abstract the 
underlying physical SDN network topology, namely VeRTIGO and CoVisor.
Moreover, as indicated in Table~\ref{tab:abstraction_comparison},
a handful of decentralized hypervisors, namely FlowN, Network
Hypervisor, AutoSlice, and OpenVirteX, can abstract the underlying
physical SDN network topology. 

We note that different hypervisors abstract the topology for different reasons.
VeRTIGO, for instance, targets the flexibility of the vSDNs and their
topologies. CoVisor intends to hide unnecessary topology information
from network applications. FlowN, on the other hand, abstracts the
topology in order to facilitate the migration of virtual resources in
a transparent manner to the vSDN tenants.

\subsubsection{Physical Node and Link Resource Abstraction}
In addition to topology abstraction, attributes of
physical nodes and links can be abstracted to be provided
within a vSDN slice to the vSDN controllers.
As outlined in Section~\ref{sec:abstraction}, SDN physical node
attributes include the CPU and flow tables,
while physical link attributes include
bandwidth, queues, and buffers.
Among the centralized hypervisors that provide topology
abstraction, there are a few that go a step further and provide also
node and links abstraction, namely VeRTIGO and CoVisor.
Other hypervisors provide node and link abstraction;
however, no topology abstraction. This is mainly due to the distributed
architecture of such hypervisors, e.g., Carrier-grade.
The distribution of the hypervisor layer tends to complicate the
processing of the entire topology to derive an abstract view of a virtual slice.

As surveyed in Section~\ref{specenhyp:sec}, hypervisors for
special network types, such as wireless or optical networks,
need to consider the unique characteristics of the
node and link resources of these special network types.
For instance, in wireless networks, the characteristics of the wireless link
resources involve several aspects of physical layer radio communication,
such as the spatial distribution of radios as well as
the radio frequencies and transmission time slots).

\subsubsection{Summary}
Overall, we observe from Table~\ref{tab:abstraction_comparison} that
only less than half of the cells have a check mark (\checkmark), i.e., provide
some abstraction of the three considered network attributes.
Most existing hypervisors have focused on abstracting only one
(or two) specific network attribute(s).
However, some hypervisor studies have not addressed abstraction at all,
e.g., FlowVisor, Double FlowVisor, Mobile FlowVisor,
Enterprise Visor, and HyperFlex.
Abstraction of the physical network attributes, i.e.,
the network topology, as well as the node and link
resources, is a complex problem, particularly for distributed
hypervisor designs.
Consequently, many SDN hypervisor studies to date have
focused on the creation of the vSDN slices and their isolation.
We expect that as hypervisors for virtualizing SDN networks mature,
more studies will strive to incorporate abstraction
mechanisms for two and possibly all three physical network attributes.

\subsection{Virtual SDN Network Isolation Attributes and Mechanisms}
\label{sumiso:sec}
\begin{center}
\begin{table*}[t]
    \centering
    \caption{Summary comparison of virtual SDN network
         isolation attributes} \label{tab:isolation_comparison}
    \begin{tabular}{|l|l|l|p{3cm}|l |p{2cm}|p{2cm}|p{3cm}| }
    \hline
      \multicolumn{3}{|l|}{} & & Control Plane
       & \multicolumn{2}{c|}{Data Plane} & Virtual SDN Network \\
\multicolumn{3}{|l|}{Classification} & Hypervisor & Instances & Nodes & Links & Addressing \\ \hline \hline
\multirow{20}{*}{Centr.} & \multirow{20}{*}
{\begin{minipage}{0.3in}Gen. Comp. Platf.\end{minipage}} &
\multirow{10}{*}
{\begin{minipage}{0.4in}Gen., Secs.~\ref{FlowVisor:sec}, \ref{cenotherhyp:sec}\end{minipage}} &
FlowVisor~\cite{Sherwood2009} & -- & Flow Tables, CPU & Bandwidth (BW) & Configurable non--overlap. flowspace\\ \cline{4-8}
& & & ADVisor~\cite{Salvadori2011} & -- & Flow Tables, CPU & BW & VLAN, MPLS  \\ \cline{4-8}
& & & VeRTIGO~\cite{Corin2012} & -- & -- & -- & -- \\ \cline{4-8}
& & & Enhanced FlowVisor~\cite{Min2012} & -- & --  & BW &-- \\ \cline{4-8}
& & & Slice Isolator~\cite{El-Azzab2011} & -- & Flow Tables & BW, queues, buffers & --\\ \cline{4-8}
& & & Double FlowVisor~\cite{Yin2013} & -- & -- & -- & -- \\ \cline{3-8}

& & \multirow{8}{*}{\begin{minipage}{0.3in}Spec. Netw. Types, Sec.~\ref{specenhyp:sec}\end{minipage}} & CellVisor~\cite{li2012tow,Li} & -- & -- & Radio (space, time, freq.) & MAC (base station) \\ \cline{4-8}
& & & RadioVisor~\cite{Gudipati2014} & -- & -- & Radio (space, time, freq.) & -- \\ \cline{4-8}
& & & MobileVisor~\cite{Nguyen2014} & -- & -- & -- & -- \\ \cline{4-8}
& & & Optical FV~\cite{Azodolmolky2012} & -- & -- & Optical (wavelength) & --  \\ \cline{4-8}
& & & Enterprise Visor\cite{Chen2014a} & -- & -- & -- & --\\ \cline{3-8}

& & \multirow{3}{*}{\begin{minipage}{0.5in}Policy-b., Sec.~\ref{polcenhyp:sec}\end{minipage}} & Compositional Hypervisor~\cite{Rexford2014} & Rules & & & \\ \cline{4-8}
& & & CoVisor~\cite{Jin2015} & Rules & & & \\ \hline \hline

\multirow{20}{*}{Distr.} &
\multicolumn{2}{l|}{\multirow{5}{*}
 {\begin{minipage}{0.8in}Gen. Comp. Platf., Sec.~\ref{gcpdishyp:sec}
            \end{minipage}}} &
FlowN~\cite{Drutskoy2013b} & Threads & -- & -- & VLAN \\ \cline{4-8}
& \multicolumn{2}{l|}{} & Network Hypervisor~\cite{hua2013net} & APIs & -- & -- & -- \\ \cline{4-8}
& \multicolumn{2}{l|}{} & AutoSlice~\cite{Bozakov2012,Bozakov2013} & -- & Flow Tables && VLAN, MPLS \\ \cline{4-8}
& \multicolumn{2}{l|}{} & Netw. Virt. Platform (NVP)~\cite{Koponen2014} & -- & Flow Tables & BW, queues & -- \\ \cline{2-8}

& \multicolumn{2}{l|}{\multirow{4}{*}{\begin{minipage}{0.8in}Gen. Comp. Platf. + Gen.-purp. NE, Sec.~\ref{gcpspne:sec} \end{minipage}}} & OpenVirteX~\cite{Al-Shabibi2014,Berde2014, Al-shabibi2014b} & -- & Flow Tables, CPU & BW & Configurable overlapping flowspace \\ \cline{4-8}
& \multicolumn{2}{l|}{} & OF NV Cloud~\cite{Matias2011}  & -- & Flow Tables & -- & MAC \\ \cline{4-8}
& \multicolumn{2}{l|}{} & AutoVFlow~\cite{Yamanaka2014} & -- & -- & -- & -- \\ \cline{2-8}

& \multicolumn{2}{l|}{\multirow{7}{*}{\begin{minipage}{0.8in}Gen. Comp. Platf. + Spec.-purp. NE, Sec.~\ref{gcpspne:sec}\end{minipage}}} &Carrier--grade~\cite{Devlic2012a} & -- & Flow tables & BW & VLAN, VxLAN, PWE  \\ \cline{4-8}
& \multicolumn{2}{l|}{} & Datapath Centric~\cite{DoriguzziCorin2014} & -- & -- & -- & -- \\ \cline{4-8}
& \multicolumn{2}{l|}{} & DFVisor~\cite{Liao2014,Liao2015} & -- & -- & -- & GRE \\ \cline{4-8}
& \multicolumn{2}{l|}{} & OpenSlice~\cite{liu2013os} & -- & -- & Optical (wavel., time)& -- \\ \cline{4-8}
& \multicolumn{2}{l|}{} & Advanced Capabilities~\cite{Sonkoly2012b} & -- & -- & BW, queues & --  \\ \cline{2-8}

& \multicolumn{2}{l|}{\begin{minipage}{0.8in}Gen. Comp. Platf. + Gen.-purp. NE + Spec.-purp. NE, Sec.~\ref{gcpgspne:sec}\end{minipage}} & Hyperflex~\cite{ble2015hyp} & CPU & -- & -- & -- \\ \hline   \hline
  \end{tabular}
\end{table*}
\end{center}
Another differentiation between the SDN hypervisors is in terms of the
isolation capabilities that they can provide between the different
vSDN slices, as summarized in Table~\ref{tab:isolation_comparison}.
As outlined in Section~\ref{sec:isolation}, there are three main network
attributes that require isolation, namely the
hypervisor resources in the control plane, the physical nodes and
links in the data plane, and the addressing of the vSDN slices.

\subsubsection{Control Plane Isolation}
Control place isolation should encompass the hypervisor instances
as well as the network communication used for the hypervisor functions,
as indicated in Fig.~\ref{fig:iso_tree}. However, existing hypervisor designs
have only addressed the isolation of the instances, and we limit
Table~\ref{tab:isolation_comparison} and the following discussion
therefore to the isolation of hypervisor instances.
Only few existing hypervisors can isolate the hypervisor
instances in the control plane, e.g., Compositional Hypervisor,
FlowN, and HyperFlex.
Compositional Hypervisor and CoVisor isolate the control plane instances
in terms of network rules and policies. Both
hypervisors compose the rules enforced by vSDN
controllers to maintain consistent infrastructure control. They also
ensure a consistent and non-conflicting state for the SDN physical
infrastructure.

FlowN isolates the different hypervisor instances (slices) in terms of the
process handling by allocating a processing thread to each slice.
The allocation of multiple processing threads avoids the blocking and
interference that would occur if the control planes of
multiple slices were to share a single thread.
Alternatively, Network Hypervisor provides isolation in terms of the APIs
used to interact with underlying physical SDN infrastructures.

HyperFlex ensures hypervisor resource isolation in terms of CPU by
restraining the control traffic of each slice. HyperFlex restrains the
slice control traffic either by limiting the control traffic at the
hypervisor software or by throttling the control traffic throughput on
the network.

\subsubsection{Data Plane Isolation}
Several hypervisors have aimed at isolating the physical
data plane resources. For instance, considering the data plane SDN physical
nodes, FlowVisor and ADVisor split the SDN flow tables and
assign CPU resources.
The CPU resources are indirectly assigned by controlling the OF
control messages of each slice.

Several hypervisors isolate the link resources in the data plane.
For instance, Slices Isolator and Advanced Capabilities,
can provide bandwidth guarantees and separate queues or buffers to
each virtual slice.
Domain/Technology-specific hypervisors, e.g.,
RadioVisor and OpenSlice, focus on providing link isolation in their
respective domains. For instance, RadioVisor splits the radio link resources
according to the frequency, time, and space dimensions,
while OpenSlice adapts isolation to the optical link resources, i.e., the
wavelength and time dimensions.

\subsubsection{vSDN Address Isolation}   \label{vsdnaddriso:sec}
Finally, isolation is needed for the addressing of vSDN slices, i.e.,
the unique identification of each vSDN (and its corresponding tenant).
There have been different mechanisms for providing virtual identifiers for
SDN slices. FlowVisor has the flexibility to assign any of the OF fields as
an identifier for the virtual slices.
However, FlowVisor has to
ensure that the field is used by all virtual slices. For instance, if
the VLAN PCP field (see Section~\ref{absisoftfv:sec})
is used as the identifier for the virtual slices, then all
tenants have to use the VLAN PCP field as slice identifier
and cannot use the VLAN PCP field otherwise in their slice.
Consequently, the
flowspace of all tenants is reduced by the identification header.
This in turn ensures that the flowspaces of all tenants are non-overlapping.

Several proposals have defined specific addressing header fields as
slice identifier,
e.g., ADVisor the VLAN tag, AutoSlice the MPLS labels and VLAN tags,
and DFVisor the GRE labels.
These different fields have generally different matching performance
in OF switches.
The studies on these hypervisors have
attempted to determine the identification header for virtual
slices that would not reduce the performance of today's OF
switches. However, the limitation of not providing the vSDN
controllers with the full flowspace, to match on for their clients, still
persists.

OpenVirteX has attempted to solve this
flowspace limitation problem. The solution proposed by OpenVirteX is
to rewrite the packets at the edge of the physical network by virtual
identifiers that are locally known by OpenVirteX.
 Each tenant can use the entire flowspace.
Thus, the virtual identification fields can be flexibly changed by OpenVirteX.
OpenVirteX transparently rewrites the flowspace
with virtual addresses, achieving a configurable overlapping
flowspace for the virtual SDN slices.

\subsubsection{Summary}
Our comparison, as summarized in Table~\ref{tab:isolation_comparison},
indicates that a lesson learned from our survey is that none of the
existing hypervisors fully isolates all network attributes.
Most hypervisor designs focus on the isolation of one or two
network attributes. Relatively few hypervisors isolate three of the
considered network attributes, namely FlowVisor, OpenVirteX, and
Carrier-grade isolate node and link resources on the data plane, as well
as the vSDN addressing space.
We believe that comprehensively addressing the isolation of
network attributes is an important direction for future research.

\section{Hypervisor Performance Evaluation Framework}
\label{perf:sec}
In this section, we outline a novel performance evaluation framework
for virtualization hypervisors for SDN networks.
More specifically, we outline the performance evaluation of pure
virtualized SDN environments.
In a pure virtualized SDN environment, the hypervisor has full control
over the entire physical network infrastructure.
(We do not consider the operation of a hypervisor in parallel with
legacy SDN networks, since available hypervisors do not support such
a parallel operation scenario.)
As background on the performance evaluation of virtual SDN networks, we
first provide an overview of existing benchmarking tools and their use in
non-virtualized SDN networks.
In particular, we summarize the state-of-the-art
benchmarking tools and performance metrics for
SDN switches, i.e., for evaluating the data plane performance,
and for SDN controllers, i.e., for evaluating the control plane performance.
Subsequently, we introduce our performance evaluation framework for
benchmarking hypervisors for creating virtual SDN networks (vSDNs).
We outline SDN hypervisor performance metrics that are derived from the
traditional performance metrics for non-virtualized environments.
In addition, based on virtual network embedding (VNE) performance metrics
(see Section~\ref{relwork:secA}), we outline general performance metrics for
virtual SDN environments.

\subsection{Background: Benchmarking Non-virtualized SDN Networks}
\label{bmnonvsdn:sec}
In this subsection, we present existing benchmarking tools and performance
metrics for non-virtualized SDN networks.
In a non-virtualized (conventional) SDN network,
there is no hypervisor; instead,
a single SDN controller directly interacts with the network of physical
SDN switches, as illustrated in Fig~\ref{concepts:fig}(a).

\subsubsection{SDN Switch Benchmarking}
\label{SDNswitchbm:sec}
\begin{figure*}[!tb]
    \centering
    \subfigure[Non-virtualized SDN switch benchmarking]{
        \includegraphics[width=0.31\textwidth]{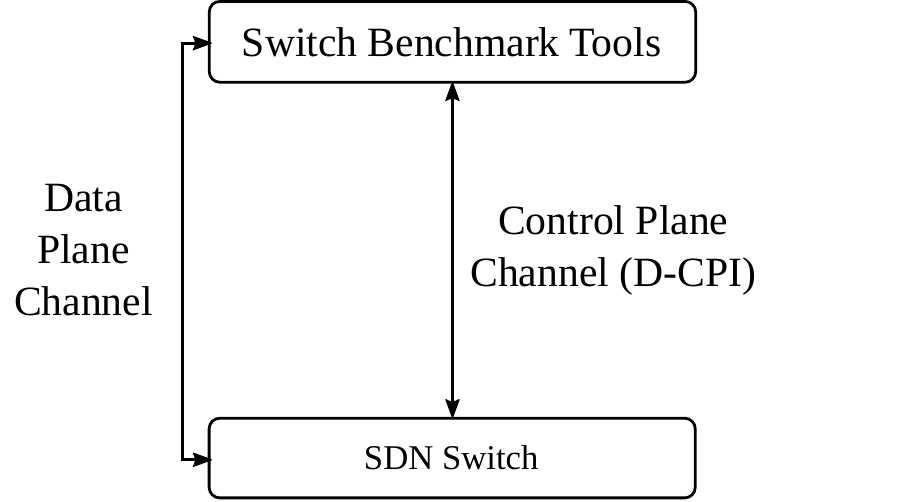}%
        \label{fig:sw_be}
    }
\subfigure[Virtualized SDN switch benchmarking: single vSDN switch]{
    \includegraphics[width=0.31\textwidth]{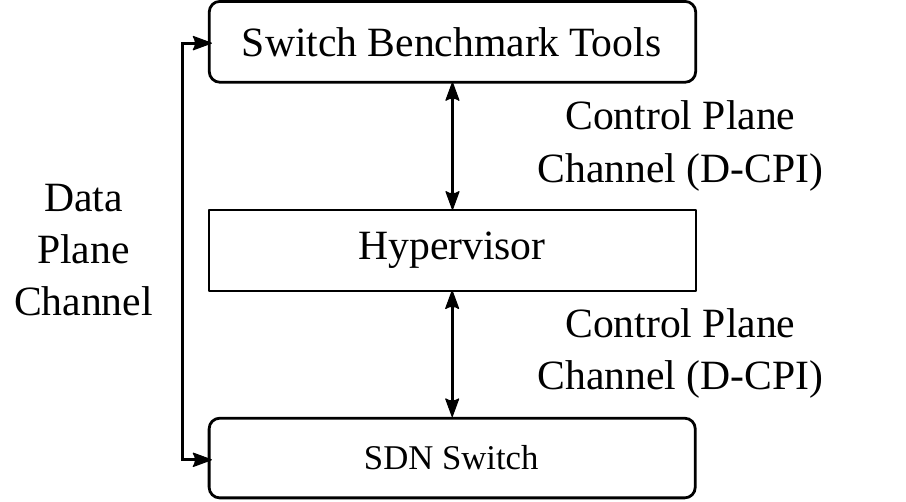}%
        \label{fig:sw_be_hy}
    }
\subfigure[Virtualized SDN switch benchmarking: multiple vSDN switches]{
	\includegraphics[width=0.31\textwidth]{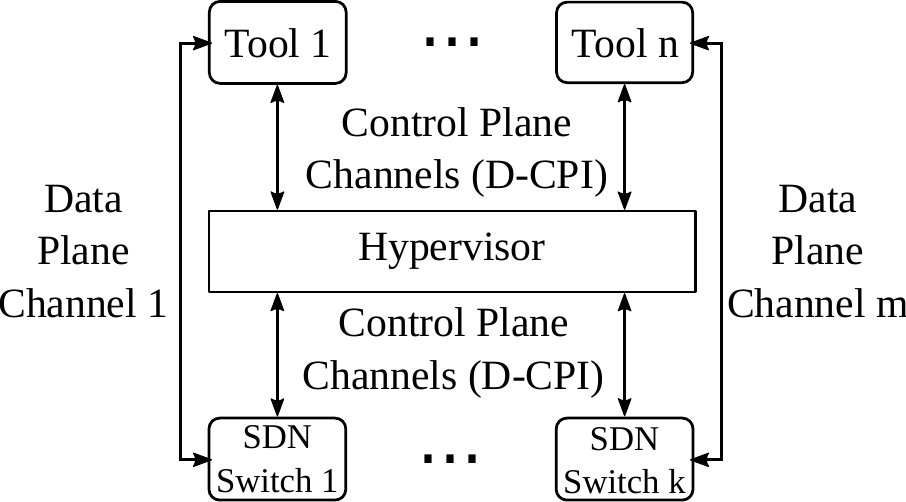}%
        \label{fig:nsw_be_hy}
    }
    \caption{Conceptual comparison of switch benchmarking set-ups:
    Switch benchmark (evaluation) tools are directly connected to
  an SDN switch in the non-virtualized SDN set-up. In the virtualized
   SDN set-ups, the control plane channel (D-CPI) passes through
   the evaluated hypervisor.}
    \label{fig:swbench}
\end{figure*}

In a non-virtualized SDN network, the SDN switch benchmark tool
directly measures the performance of one SDN switch, as illustrated in
Fig.~\ref{fig:swbench}(a).
Accordingly, switch benchmarking tools play two roles:
The first role is that of the SDN controller as illustrated in
Figure~\ref{concepts:fig}(a). The switch benchmarking tools connect to the
SDN switches via the control plane channels (D-CPI).
In the role of SDN controller, the tools can, for instance,
measure the time delay between sending OF status requests and
receiving the reply under varying load scenarios.
Switch benchmarking tools are also connecting to the data plane of switches.
This enables the tools to play their second role, namely the measurement of the
entire chain of processing elements of network packets
Specifically, the tools can measure
the time from sending a packet into the data plane,
to being processed by the controller, and finally to being forwarded on the
data plane.

Examples of SDN switch benchmarking tools are OFTest~\cite{OFTest},
OFLOPS~\cite{rot2012ofl}, and FLOPS-Turbo~\cite{rot2014ope}.
OFTest was developed to verify the correct implementation of the OF
protocol specifications of SDN switches. OFLOPS was developed in
order to shed light on the different implementation details of SDN
switches. Although the OF protocol provides a common interface to SDN switches,
implementation details of SDN switches are vendor-specific. Thus, for
different OF operations, switches from different vendors may exhibit
varying performance. The OFLOPS SDN switch
measurements target the OF packet processing actions as well as the update
rate of the OF flow table and the resulting impact on the data plane
performance.
OFLOPS can also evaluate the monitoring provided by OF and cross-effects of
different OF operations. In the following, we explain these
different performance metrics in more detail:

\paragraph{OF Packet Processing} OF packet processing encompasses all
operations that directly handle network packets,
e.g., matching, forwarding, or dropping.
Further, the current OF specification defines
several packet modifications actions. These packet
modification actions can rewrite protocol header fields,
such as MAC and IP addresses, as well as protocol-specific VLAN fields.

\paragraph{OF Flow Table Updates} OF flow table updates add, delete, or
update existing flow table entries in an SDN switch. Flow table
updates result from control decisions made by the SDN
controller. As flow tables play a key role in SDN networks
(comparable to the Forwarding Information Base (FIB)~\cite{tro2001ter}
in legacy networks), SDN switch updates should be
completed within hundreds of microseconds
as in legacy networks~\cite{sha2001exp}.
OFLOPS measures the time until an action has been applied on the switch.
A possible measurement approach uses barrier request/reply messages,
which are defined by the OF specification~\cite{OF14,OF15}.
After sending an OF message, e.g., a flow mod message, the switch
replies with a barrier reply message when the
flow mod request has actually been processed. As it was shown
in~\cite{Epfl-report-} that switches may send barrier reply messages
even before a rule has really been applied, a second measurement approach is
needed.
The second approach incorporates the effects on the data plane by
measuring the time period from the instant when a rule was sent to the instant
when its effect can be recognized on the data plane. The effect can, for
instance, be a packet that is forwarded after a forwarding rule has been added.

\paragraph{OF Monitoring Capabilities}
OF monitoring capabilities provide flow statistics.
Besides per-flow statistics, OF can provide
statistics about aggregates, i.e., aggregated flows. Statistics count
bytes and packets per flow or per flow aggregate.
The current statistics can serve as basis for network traffic
estimation and monitoring~\cite{too2010ope,yu2013flo,van2014ope}
as well as dynamic adoptions by SDN applications and traffic
engineering~\cite{aga2013tra,aky2014roa}.
Accurate up-to-date statistical information is therefore important.
Accordingly, the time to receive
statistics and the consistency of the statistical information are
performance metrics for the monitoring capabilities.

\paragraph{OF Operations Cross-Effects and Impacts}
\label{sdn_ofcross:sec}
All available OF features can be used simultaneously while operating SDN
networks. For instance, while an SDN switch makes traffic steering decisions,
an SDN controller may request current statistics.
With network virtualization, which is considered in detail in
Section~\ref{benchvir:sec}, a mixed operation of SDN switches
(i.e., working simultaneously on a mix of OF features) is common.
Accordingly, measuring the performance of an SDN
switch and network under varying OF usage scenarios is important,
especially for vSDNs.

\paragraph{Data Plane Throughput and Processing Delay}
\label{legacy_throu_proc:sec}
In addition to OF-specific performance metrics for SDN switches, legacy
performance metrics should also be applied to evaluate the performance
of a switch entity. These legacy performance metrics are mainly the data plane
throughput and processing delay for a specific switch task. The data plane
throughput is usually defined as the rate of packets (or bits) per second
that a switch can forward.
The evaluation should include a range of additional
tasks, e.g., labeling or VLAN tagging. Although legacy switches are
assumed to provide line rate throughput for simple tasks, e.g., layer~2
forwarding, they may exhibit varying performance for more complex tasks,
e.g., labeling tasks. The switch processing time typically depends on
the complexity of a specific task and is expressed in terms of the time
that the switch needs to complete a specific single operation.

\subsubsection{SDN Controller Benchmarking}
\label{sdncontbm:sec}
\begin{figure*}[!tb]
    \centering
    \subfigure[Non-virtualized SDN controller benchmarking]{
\includegraphics[width=0.31\textwidth]{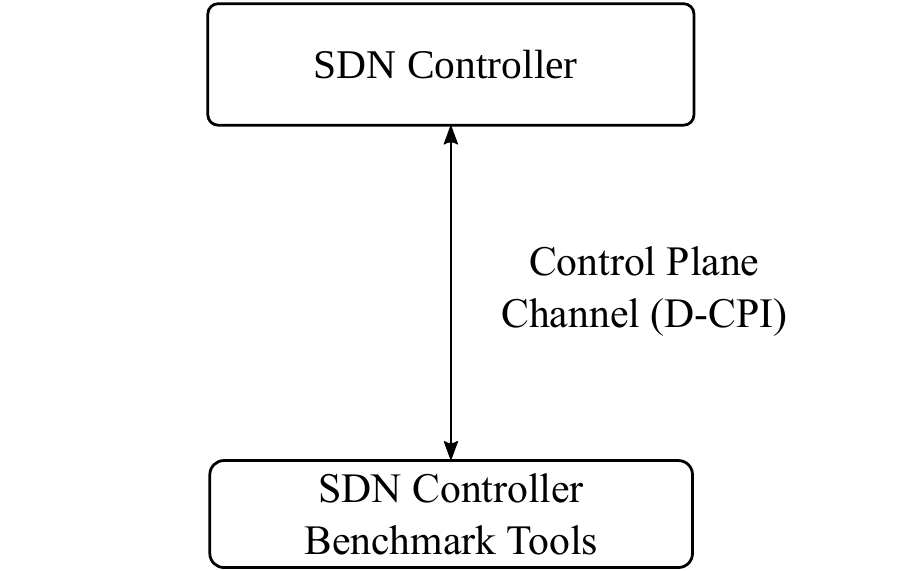}%
    \label{fig:ctrl_be}
    }
\subfigure[Virtualized SDN controller benchmarking: single vSDN controller]{
\includegraphics[width=0.31\textwidth]{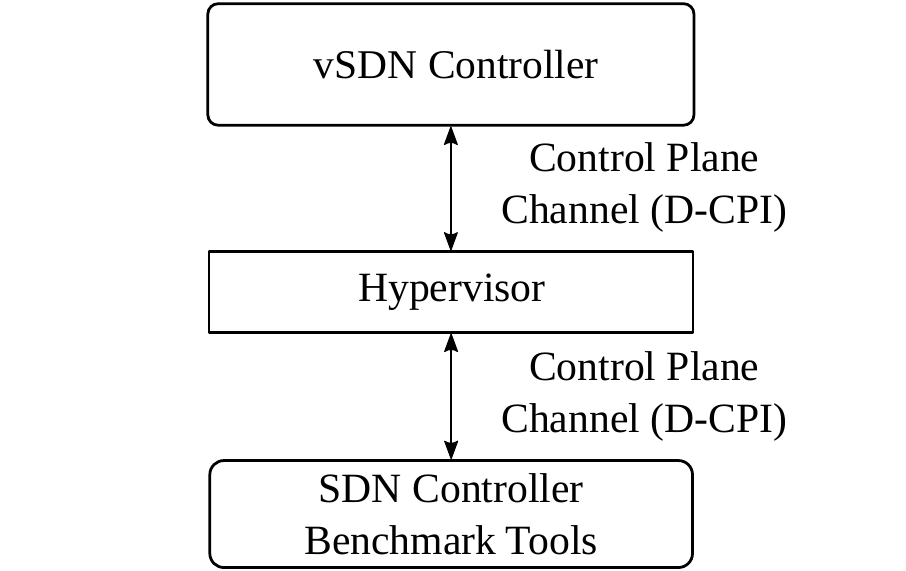}%
        \label{fig:ctrl_be_hy}
    }
\subfigure[Virtualized SDN controller benchmarking: multiple vSDN controllers]{
\includegraphics[width=0.31\textwidth]{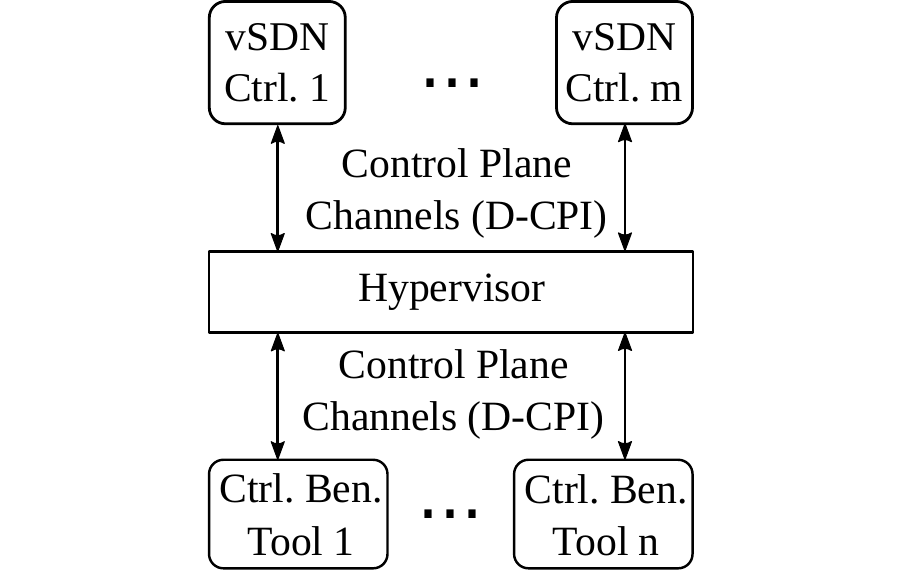}%
        \label{fig:nctrl_be_hy}
    }
    \caption{Conceptual comparison of controller benchmarking set-ups:
In the non-virtualized set-up, the SDN controller benchmark (evaluation)
tools are directly connected to the SDN controller. In the vSDN set-ups, the
control plane channel (D-CPI) traverses the evaluated hypervisor.}
    \label{fig:ctrlbench}
\end{figure*}

Implemented in software, SDN controllers can have a significant impact
on the performance of SDN networks. Fig.~\ref{fig:ctrl_be}
shows the measurement setup for controller benchmarks.
The controller benchmark tool emulates SDN switches, i.e., the benchmark tool
plays the role of the physical SDN network (the network of SDN switches)
in Fig.~\ref{concepts:fig}(a).
The benchmark tool can send arbitrary OF requests to the SDN controller.
Further, it can emulate an arbitrary number of switches,
e.g., to examine the scalability of an SDN controller.
Existing SDN controller benchmark tools include Cbench~\cite{ToGGC12},
OFCBenchmark~\cite{jar2012fle}, OFCProbe~\cite{Jarschel2014a}, and
PktBlaster~\cite{pktblaster}.
In~\cite{ToGGC12}, two performance
metrics of SDN controllers were defined, namely the controller OF message
throughput and the controller response time.
\paragraph{Controller OF Message Throughput} The OF message
throughput specifies the rate of messages (in units of messages/s)
that an SDN controller can process on average. This throughput is an
indicator for the scalability of SDN controllers~\cite{yeg2013sca}.
In large-scale networks, SDN controllers may have to serve on the order of
thousands of  switches. Thus, a high controller throughout of OF messages
is highly important.
For instance, for each new connection, a switch sends a OF PCKT\_IN
 message to the controller.  When many new flow connections are
requested, the controller should respond quickly to ensure short
flow set-up times.

\paragraph{Controller Response Time} The response time of an SDN
controller is the time that the SDN controller needs to respond to a message,
e.g., the time needed to create a reply to a PCKT\_IN message.
The response time is typically related to the OF message throughput in that
a controller with a high throughput has usually a short response time.

The response time and the OF message throughput are also used as
scalability indicators. Based on the performance, it has to be decided
whether additional controllers are necessary. Furthermore, the metrics
should be used to evaluate the performance in a best-effort scenario,
with only one type of OF message, and in mixed scenarios, with
multiple simultaneous types of OF messages~\cite{Jarschel2014a}.
Evaluations and corresponding refinements of non-virtualized SDN
controllers are examined in ongoing research, see e.g.,~\cite{Shalimov2013}.
Further evaluation techniques for non-virtualized
SDN controllers have recently been studied,
e.g., the verification of the correctness of SDN controller
programs~\cite{bal2014ver,khu2012ver,kuz2012of} and the forwarding
tables~\cite{mai2011deb,zen2014lib}.

\subsection{Benchmarking SDN Network Hypervisors}
\label{benchvir:sec}
In this section, we outline a comprehensive evaluation framework for
hypervisors that create vSDNs.  We explain the range of
measurement set-ups for evaluating (benchmarking) the performance of
the hypervisor.
In general, the operation of the isolated vSDN slices should efficiently
utilize the underlying physical network.
With respect to the run-time performance, a vSDN should achieve a
performance level close to the performance level of its non-virtualized SDN
network counterpart (without a hypervisor).

\subsubsection{vSDN Embedding Considerations}
As explained in Section~\ref{relwork:secA}, with network
  virtualization, multiple virtual networks operate simultaneously
  on the same physical infrastructure. This simultaneous operation
 generally applies also to
  virtual SDN environments. That is, multiple vSDNs share the same
  physical SDN network resources (infrastructure).
In order to achieve a high utilization of the
  physical infrastructure, resource management algorithms for virtual
  resources, i.e., node and link resources, have to be applied. In
  contrast to the traditional assignment problem, i.e., the Virtual Network
  Embedding (VNE) problem (see Section~\ref{relwork:secA}),
the intrinsic attributes of an SDN environment
  have to be taken into account for vSDN embedding.
For instance, as the efficient
  operation of SDN data plane elements relies on the use of the TCAM
  space, the assignment of TCAM space has to be taken into account
  when embedding vSDNs. Accordingly, VNE algorithms, which already consider link resources, such as
  data rate, and node resources, such as CPU, have to be extended to SDN
  environments~\cite{Guerzoni,Demirci2014,Riggio2013}. The
  performance of vSDN embedding algorithms can be evaluated with
 traditional VNE metrics, such as acceptance rate or revenue/cost per vSDN.

\subsubsection{Two-step Hypervisor Benchmarking Framework}
In order to evaluate and compare SDN hypervisor performance, a general
benchmarking procedure is needed.  We outline a
two-step hypervisor benchmarking framework. The first step is to
benchmark the \textit{hypervisor system as a whole}.
This first step is needed to
quantify the hypervisor system for general use cases and set-ups, and
to identify general performance issues.  In order to reveal the
performance of the hypervisor system as a whole, explicit measurement cases
with a combination of one or multiple vSDN switches and one or
multiple vSDN controllers should be conducted.  This measurement setup
reveals the performance of the interplay of the individual
hypervisor functions.  Furthermore, it demonstrates the performance
for varying vSDN controllers and set-ups.  For example, when
benchmarking the capabilities for isolating the data plane resources,
measurements with one vSDN switch as well as multiple vSDN
switches should be conducted.

The second step is to benchmark each
\textit{specific hypervisor function} in detail.  The specific hypervisor
function benchmarking reveals bottlenecks in the
hypervisor implementation.  This is needed to compare the
implementation details of individual virtualization functions.
If hypervisors exhibit different performance levels for different
functions, an operator can select the best hypervisor for the
specific use case.  For example, a networking scenario where no
data plane isolation is needed does not require a high-performance data plane
isolation function.  In general, all OF related and conventional
metrics from Section~\ref{bmnonvsdn:sec} can be
applied to evaluate the hypervisor performance.

In the following Sections~\ref{vsdnswbm:sec} and~\ref{vsdncontbm:sec},
we explain how different vSDN switch scenarios and vSDN controller scenarios
should be evaluated. The purpose of the performance evaluation of the
vSDN switches and vSDN controllers is to draw conclusions about the
performance of the hypervisor system as a whole.
In Section~\ref{bmhvind:sec}, we outline how and why to benchmark
each individual hypervisor function in detail. Only such a comprehensive
hypervisor benchmark process can identify the utility of a hypervisor
for specific use cases or a range of use cases.

\subsubsection{vSDN Switch Benchmarking} \label{vsdnswbm:sec}
When benchmarking a hypervisor, measurements with one vSDN switch
should be performed first.
That is, the first measurements should be conducted with a
single benchmark entity and a single vSDN switch,
as illustrated in Fig.~\ref{fig:sw_be_hy}.
Comparisons of the evaluation results for the legacy SDN switch benchmarks
(for non-virtualized SDNs, see Section~\ref{SDNswitchbm:sec})
for the single vSDN switch (Fig.~\ref{fig:sw_be_hy}) with results
for the corresponding non-virtualized SDN switch (without
hypervisor, Fig.~\ref{fig:sw_be}) allows for an analysis of the
overhead that is introduced by the hypervisor.
For this single vSDN switch evaluation, all SDN switch metrics from
Section~\ref{SDNswitchbm:sec} can be employed.
A completely transparent hypervisor would show zero
overhead, e.g., no additional latency for OF flow table updates.

In contrast to the single vSDN switch set-up in Fig.~\ref{fig:sw_be_hy}, the
multi-tenancy case in Fig.~\ref{fig:nsw_be_hy} considers multiple {vSDNs}.
The switch benchmarking tools, one for every emulated vSDN controller,
are connected with the hypervisor, while
multiple switches are benchmarked as representatives for vSDN networks.
Each tool may conduct a different switch performance
measurement. The results of each measurement should be compared to the
set-up with a single vSDN switch (and a single tool) illustrated
in Fig.~\ref{fig:sw_be_hy} for each
individual tool.  Deviations from the single vSDN switch scenario may indicate
 cross-effects of the hypervisor. Furthermore,
different combinations of measurement scenarios may reveal specific
implementation issues of the hypervisor.

\subsubsection{vSDN Controller Benchmarking}  \label{vsdncontbm:sec}
Similar to vSDN switch benchmarking, the legacy SDN controller
benchmark tools (see Section~\ref{sdncontbm:sec}), should be used for a first
basic evaluation.
Again, comparing scenarios where a hypervisor is activated (vSDN scenario)
or deactivated (non-virtualized SDN scenario), allows to
draw conclusion about the overhead introduced by the hypervisor.
For the non-virtualized SDN (hypervisor deactivated) scenario,
the benchmark suite is directly connected to the SDN controller,
    as illustrated in Fig.~\ref{fig:ctrl_be}.
For the vSDN (hypervisor activated) scenario, the hypervisor
is inserted between the benchmark suite and the SDN controller,
see Fig.~\ref{fig:ctrl_be_hy}.

In contrast to the single controller set-up in Fig.~\ref{fig:ctrl_be_hy},
multiple controllers are connected to the hypervisor in the multi-tenancy
scenario in Fig.~\ref{fig:nctrl_be_hy}.
In the multi-tenancy scenario, multiple controller benchmark tools can be
used to create different vSDN topologies, whereby each
tool can conduct a different controller performance measurement.
Each single measurement should be compared to
the single virtual controller benchmark (Fig.~\ref{fig:ctrl_be_hy}).
Such comparisons quantify not only the overhead introduced by the hypervisor,
but also reveal cross-effects, that are the result of multiple
simultaneously running controller benchmark tools.

Compared to the single vSDN switch
scenario, there are two main reasons for additional overheads in the
multi-tenancy scenario. First, each virtualizing function
introduces a small amount of processing overhead.
Second, as the processing of multiple vSDNs is shared
among the hypervisor resources, different types of OF messages sent by
the vSDNs may lead to varying performance when compared to a single
vSDN set-up.  This second effect is comparable to the OF operations
cross-effect for non-virtualized SDN environments, see
Section~\ref{sdn_ofcross:sec}.  For instance, translating messages
from different vSDNs appearing at the same time may result in message
contention and delays, leading to higher and variable
processing times.  Accordingly, multiple vSDN switches should be
simultaneously considered in the performance evaluations.

\subsubsection{Benchmarking of Individual Hypervisor Functions}
\label{bmhvind:sec}
Besides applying basic vSDN switch and vSDN controller benchmarks, a
detailed hypervisor evaluation should also include the benchmarking of
the individual hypervisor functions, i.e., the abstraction and
isolation functions summarized in Figs.~\ref{fig:lev_virt}
and~\ref{fig:iso_tree}.
All OF related and legacy metrics from Section~\ref{bmnonvsdn:sec}
should be applied to evaluate the individual hypervisor functions.
The execution of the hypervisor functions may require network resources,
and thus reduce the available resources for the data plane operations of the
vSDN networks.

\paragraph{Abstraction Benchmarking}
The three main aspects of abstraction summarized in
Fig.~\ref{fig:lev_virt} should be benchmarked, namely
topology abstraction, node resource abstraction, and link resource abstraction.
Different set-ups should be considered for evaluating the topology abstraction.
One set-up is the $N$-to-1 mapping, i.e., the capability of
hypervisor to map multiple vSDN switches to
a single physical SDN switch. 
For this $N$-to-1 mapping, the performance of the non-virtualized
physical switch should be compared with the performance of
the virtualized switches. This comparison allows conclusions
about the overhead introduced by the virtualization mechanism.
For the 1-to-$N$ mapping, i.e., the mapping of 
one vSDN switch to $N$ physical SDN switches, the same procedure should be
applied. For example, presenting two physical SDN switches as one large
vSDN switch demands specific abstraction mechanisms.
The performance of the large vSDN switch should be compared to
the aggregated performance of the two physical SDN switches.

In order to implement node resource abstraction and link resource
abstraction, details about node and link capabilities of
switches have to be hidden from the vSDN controllers.
Hypervisor mapping mechanisms have to remove (filter) detailed node
and link information,
e.g., remove details from switch performance monitoring information,
at runtime. For example, when the switch reports
statistics about current CPU and memory utilization, providing only the
CPU information to the vSDN controllers
requires that the hypervisor removes the memory
information. The computational processing required for this information
filtering may degrade performance, e.g., reduce the
total OF message throughput of the hypervisor.

\paragraph{Isolation Benchmarking}
Based on our classification of isolated network attributes summarized in
Fig.~\ref{fig:iso_tree}, control plane isolation, data plane isolation,
and vSDN addressing need to be benchmarked.
The evaluation of the control plane isolation refers to the evaluation of
the resources that are involved in ensuring the isolation of the
processing and transmission of control plane messages.
Isolation mechanisms for the range of control plane resources,
including bandwidth isolation on the control channel, should be examined.
The isolation performance should be benchmarked
for a wide variety of scenarios ranging from a single to multiple
vSDN controllers that utilize or over-utilize their assigned resources.

Hypervisors that execute hypervisor functions on the data plane,
i.e., the distributed hypervisors that involve NEs
(see Sections~\ref{gcpgpne:sec}--\ref{gcpgspne:sec}),
require special consideration when evaluating control plane isolation.
The hypervisor functions may share resources with data plane
functions; thus, the isolation of the resources
has be evaluated under varying data plane workloads.

Data plane isolation encompasses the isolation of the node resources
and the isolation of the link resources.
For the evaluation of the CPU isolation, the processing time of more than
two vSDN switches should be benchmarked.
If a hypervisor implements bandwidth isolation on the
data plane, the bandwidth isolation can be benchmarked in
terms of accuracy.  If one vSDN is trying to over-utilize
its available bandwidth, no cross-effect should be seen for other vSDNs.

Different vSDN addressing schemes may show trade-offs in terms of available
addressing space and performance.
For instance, when comparing hypervisors
(see Table~\ref{tab:isolation_comparison}), there are
``matching only'' hypervisors
that define a certain part of the flowspace to be used for vSDN (slice)
identification, e.g., FlowVisor.
The users of a given slice have to use the prescribed value of their
respective tenant which the hypervisor uses for matching.
The ``add labels'' hypervisors assign a label for slice identification, e.g.,
Carrier-grade assigns MPLS labels to virtual slices,
which adds more degrees of freedom and increases the flowspace.
However, the hypervisor has to add and remove the labels before
forwarding to the next hop in the physical SDN network.
Both types of hypervisors provide a limited addressing flowspace to
the tenants. However, ``matching only'' hypervisors may perform
better than ``add labels'' hypervisors since
matching only is typically simpler for switches than matching and labeling.
Another hypervisor type modifies
the assigned virtual ID to offer an even larger flowspace, e.g.,
OpenVirteX rewrites a set of the MAC or IP header bits at the edge
switches of the physical SDN network. This can also impact the hypervisor
performance as the SDN switches need to ``match and modify''.

Accordingly, when evaluating different
addressing schemes, the size of the addressing space has
to be taken into account when interpreting performance.
The actual overhead due to an addressing scheme
can be measured in terms of the introduced processing overhead.
Besides processing overhead, also the
resulting throughput of different schemes can be measured, or
formally analyzed.

\section{Future Research Directions}\label{fut:sec}
From the preceding survey sections, we can draw several observations
and lessons learned that indicate future research directions.
First, we can observe the vast heterogeneity
and variety of SDN hypervisors.
There is also a large set of network attributes of the data and
control planes that need to be selected for abstraction and isolation.
The combination of a particular hypervisor along with selected network
attributes for abstraction and isolation has a direct impact on the resulting
network performance, whereby different trade-offs result
from different combinations of selections.
One main lesson learnt from the survey is that the studies completed
so far have demonstrated the concept of SDN hypervisors as being
feasible and have led to general (albeit vague) lessons learnt mainly
for abstraction and isolation, as summarized in Section~\ref{sec:summary}.
However, our survey revealed a pronounced lack of rigorous, comprehensive
performance evaluations and comparisons of SDN hypervisors.
In particular, our survey suggests that there is currently no single best or
simplest hypervisor design.
Rather, there are many open research questions
in the area of vSDN hypervisors and a pronounced need for
detailed comprehensive vSDN hypervisor performance evaluations.

In this section, we outline future research directions. While we address
open questions arising from existing hypervisor developments, we also
point out how hypervisors can advance existing non-virtualized SDN networks.

\subsection{Service Level Definition and Agreement for vSDNs}
In SDN networks, the control plane performance can directly affect
the performance of the data plane. Accordingly, research on conventional
SDN networks identifies and optimizes SDN control planes. Based
on the control plane demands in SDN networks, vSDN tenants may also
need to specify their OF-specific demands. Accordingly, in addition to
requesting virtual network topologies in terms of virtual node demands
and virtual link demands, tenants may need to specify
control plane demands. Control plane demands include control message
throughput or latency, which may differ for different control messages
defined by the control API. Furthermore, it may be necessary to define
demands for specific OF message types. For example, a service may not
demand a reliable OF stats procedure, but may require fast OF flow-mod
message processing. A standardized way for defining these
demands and expressing them in terms of OF-specific parameters is not yet
available. In particular, for vSDNs, such a specification is
needed in the future.

\subsection{Bootstrapping vSDNs}
The core motivation for introducing SDN is programmability and
  automation. However a clear bootstrapping procedure for vSDNs is
missing. Presently, the bootstrapping of vSDNs does not follow a pre-defined
mechanisms. In general, before a vSDN is ready
for use,  all involved components should be bootstrapped and
connected.
More specifically, the connections between the
  hypervisor and controllers needs to established. The virtual slices in
  the SDN physical network need to be instantiated. Hypervisor
  resources need to be assigned to slices. Otherwise, problems may arise,
e.g., vSDN switches may already send OF PCKT\_IN
messages even though the vSDN controller is still bootstrapping. In
case of multiple uncontrolled vSDNs, these messages may even
overload the hypervisor. Clear
bootstrapping procedures are important, in particular, for fast and
dynamic vSDN creation and setup. Missing bootstrapping
procedures can even lead to undefined states,
i.e., to non-deterministic operations of the vSDN networks.

\subsection{Hypervisor Performance Benchmarks}
As we outlined in Section~\ref{perf:sec}, hypervisors need detailed
performance benchmarks. These performance benchmarks should encompass
every possible performance aspect that is considered in today's legacy
networks. For instance, hypervisors integrated in specific networking
environments, e.g., mobile networks or enterprise networks, should be
specifically benchmarked with respect to the characteristics of the
networking environment. In mobile networks, virtualization solutions for SDN
networks that interconnect access points may have to support
mobility characteristics, such as a high rate of handovers per second.
On the other hand, virtualization solutions for enterprise networks
should provide reliable and secure communications.
Hypervisors virtualizing networks that
host vSDNs serving highly dynamic traffic patterns may need to
provide fast adaptation mechanisms.

Besides benchmarking the runtime performance of hypervisors,
the efficient allocation of physical resources to {vSDNs} needs
  further consideration in the future. The general assignment problem
  of physical to virtual resources, which is commonly known as the
Virtual Network Embedding (VNE) problem~\cite{fis2013vir},
has to be further extended to virtual SDN
  networks. Although some first studies
exist~\cite{Guerzoni,Demirci2014,Riggio2013},
they neglect the impact of the
  hypervisor realization on the resources required to accept a
  vSDN. In particular, as hypervisors realize virtualization functions
  differently, the different functions may demand varying physical
  resources when accepting vSDNs.
Accordingly, the hypervisor design
  has to be taken into account during the embedding process. Although
  generic VNE algorithms for general network
  virtualization exist~\cite{fis2013vir},
there is an open research question of how to model
  different hypervisors and how to integrate them into embedding algorithms in
  order to be able to provide an efficient assignment of physical
  resources.

\subsection{Hypervisor Reliability and Fault Tolerance}
The reliability of the hypervisor layer needs to be investigated in detail as
a crucial aspect towards an actual deployment of vSDNs. First,
mechanisms should be defined to recover from hypervisor failures and
faults. A hypervisor failure can have significant impact. For example,
a vSDN controller can lose control of its vSDN, i.e., experience a
vSDN blackout, if the connection to the hypervisor is temporally lost
or terminated. Precise procedures and
mechanisms need to be defined and implemented by both hypervisors and
controllers to recover from hypervisor failures.

Second, the hypervisor
development process has to include and set up levels of redundancy
to be able to offer a reliable virtualization of SDN networks.
This redundancy adds to the
management and logical processing of the hypervisor and may degrade performance
in normal operation conditions (without any failures).

\subsection{Hardware and Hypervisor Abstraction}
As full network virtualization aims at deterministic performance
guarantees for vSDNs, more research on SDN hardware virtualization
needs to be conducted. The hardware isolation and abstraction provided by
different SDN switches can vary significantly. Several limitations and
bottlenecks can be observed, e.g., an SDN switch may not be able
to instantiate an OF agent for each virtual slice. Existing hypervisors
try to provide solutions that indirectly achieve hardware
isolation between the virtual slices. For example, FlowVisor limits
the amount of OF messages per slice in order to indirectly isolate the
switch processing capacity, i.e., switch CPU. As switches show varying
processing times for different OF messages, the FlowVisor concept
would demand a detailed a priori vSDN switch benchmarking.
Alternatively, the capability to assign hardware resources
to vSDNs would facilitate (empower) the entire virtualization
paradigm.

Although SDN promises to provide a standardized interface to switches,
existing switch diversity due to vendor variety leads to performance
variation, e.g., for QoS configurations of switches. These issues have been
identified in recent studies, such
as~\cite{laz2014tan,dur2015per,KuzniarBook,Epfl-report-,Mohan2013,Ngoc2015},
for non-virtualized SDN networks. Hypervisors may be designed to
abstract switch diversity in vendor-heterogenous
environments. Solutions, such as Tango~\cite{laz2014tan},
should be integrated into
existing hypervisor architectures. Tango provides on-line switch
performance benchmarks. SDN applications can integrate these
SDN switch performance benchmarks while making, e.g., steering decisions.
However, as research on switch diversity is still in its
infancy, the integration of existing solutions into hypervisors is an
open problem. Accordingly, future research should examine how
to integrate mechanisms that provide deterministic switch
performance models into hypervisor designs.

\subsection{Scalable Hypervisor Design}
In practical SDN virtualization deployments, a single hypervisor entity
would most likely not suffice. Hence, hypervisor designs need to
consider the distributed architectures in more detail.
Hypervisor scalability needs to be addressed by defining and
examining the operation of the
hypervisor as a whole in case of distribution. For instance, FlowN
simply sends a message from one controller server (say,
responsible for the physical switch) to another (running the tenant
controller application) over a TCP connection. More efficient
algorithms for assigning tenants and switches to hypervisors are an
interesting area for future research. 
An initial approach for dynamic (during run-time) assignment 
of virtual switches and tenant controllers 
to distributed hypervisors has been introduced in~\cite{Basta2015}.
Additionally, the hypervisors need to be developed and 
implemented with varying granularity of distribution, 
e.g., ranging from distribution of whole instances to distribution 
of modular functions.

\subsection{Hypervisor Placement}
Hypervisors are placed between tenant controllers and vSDN networks.
Physical SDN networks may have a wide geographical distribution.
Thus, similar to the controller placement problem in non-virtualized
SDN environments~\cite{Heller2012,Hu2013}, the placement of the
hypervisor demands detailed investigations.  In addition to the
physical SDN network, the hypervisor placement has to consider the
distribution of the demanded vSDN switch locations and the locations
of the tenant controllers.  If a hypervisor is implemented through
multiple distributed hypervisor functions, i.e., distributed
abstraction and isolation functions, these functions have to be
carefully placed, e.g., in an efficient hypervisor function chain.
For distributed hypervisors, the network that provides the
communications infrastructure for the hypervisor management plane has
to be taken into account.  In contrast to the SDN controller placement
problem, the hypervisor placement problems adds multiple new
dimensions, constraints, and possibilities.  
An initial study of network hypervisor placement~\cite{Blenk2015hpp} has 
provided a mathematical model for
  analyzing the placement of hypervisors when node and link constraints
  are not considered.  Similar to the initial study of the controller
  placement problem~\cite{Heller2012}, the network hypervisor
  placement solutions were optimized and analyzed with respect to
  control plane latency.
Future research on the hypervisor placement problem should also
consider that hypervisors may have to
serve dynamically changing vSDNs, giving rise to 
dynamic hypervisor placement problems.

\subsection{Hypervisors for Special Network Types}
While the majority of hypervisor designs
to date have been developed for generic wired networks,
there have been only relatively few initial studies
on hypervisor designs for special network types, such as wireless
and optical networks, as surveyed in Section~\ref{specenhyp:sec}.
However, networks of a special type, such as wireless and optical
networks, play a very important role in the Internet today.
Indeed, a large portion of the Internet traffic emanates from
or is destined to wireless mobile devices; similarly, large portions
of the Internet traffic traverse optical networks.
Hence, the development of SDN hypervisors that account for
the unique characteristics of special network types,
e.g., the characteristics of the wireless or optical
transmissions, appears to be highly important.
In wireless networks in particular, the flexibility of
offering a variety of services based a given physical
wireless network infrastructure is highly
appealing~\cite{cao2015sof,chen2015clo,che2015sel,gra2015sof,li2015edi,sun2015int}.
Similarly, in the area of access (first mile) networks, where
the investment cost of the network needs to be amortized from
the services to a relatively limited subscriber 
community~\cite{bia2013cos,kra2012evo,mcg2012inv,yan2003gen},
virtualizing the installed physical access network
is a very promising strategy~\cite{yan2014sof,ker2014sof,kon2015sdn,val2015exp,yan2013ope}.

Another example of a special network type for SDN virtualization
is the sensor network~\cite{han2014nov,jac2015usi,mou2015nfv,say2014res}.
Sensor networks have unique limitations due to the limited
resources on the sensor nodes that sense the environment and transmit
the sensing data over wireless links to gateways or sink 
nodes~\cite{fad2015sur,rein2011low,see2011tow,tav2012sur}.
Virtualization and hypervisor designs for wireless sensor networks
need to accommodate these unique characteristics.

As the underlying physical networks further evolve and new
networking techniques emerge, hypervisors implementing
the SDN network virtualization need to adapt. That is, the adaption of
hypervisor designs to newly evolving networking techniques is an ongoing
research challenge.
More broadly, future research needs to examine which hypervisor designs
are best suited for a specific network type in combination with a
particular scale (size) of the network.

\subsection{Self-configuring and Self-optimizing Hypervisors}
Hypervisors should always try to provide the best possible
virtualization performance for different network topologies,
independent of the realization of the underlying SDN networking
hardware, and for varying vSDN network demands. In order to
continuously strive for the best performance,
hypervisors may have to be designed to become
highly adaptable.  Accordingly, hypervisors should implement
mechanisms for self-configuration and self-optimization. These
operations need to work on short time-scales in order to achieve
high resource efficiency for the virtualized resources. Cognitive and
learning-based hypervisors may be needed to improve hypervisor
operations. Furthermore, the self-reconfiguration should be
transparent to the performance of the vSDNs and incur minimal
configuration overhead for the hypervisor operator. Fundamental
questions are how often and how fast a hypervisor should react to
changing vSDN demands under varying optimization objectives for the
hypervisor operator. Optimization objectives include
energy-awareness, balanced network load, and high reliability. The
design of hypervisor resource management algorithms solving these
challenges is an open research field and needs detailed investigation in future
research.

\subsection{Hypervisor Security}
OF proposes to use encrypted TCP connections between controllers and
switches. As hypervisors intercept these connections,
a hypervisor should provide a trusted encryption platform. In
particular, if vSDN customers connect to multiple different
hypervisors, as it may occur in multi-infrastructure
environments, a trusted key distribution and management system becomes
necessary. Furthermore, as secure technologies may add additional
processing overhead, different solutions need to be benchmarked for
different levels of required security. The definition of
hypervisor protection measures against attacks is required.
A hypervisor has to protect itself from attacks by defining policies
for all traffic types, including traffic that does not belong
to a defined virtual slice.

\section{Conclusion}
\label{concl:sec}
We have conducted a comprehensive survey of hypervisors for
virtualizing software defined networks (SDNs).
A hypervisor abstracts (virtualizes) the underlying physical SDN
network and allows multiple users (tenants) to share the
underlying physical SDN network.
The hypervisor slices the underlying physical SDN network into
multiple slices, i.e., multiple virtual SDN networks (vSDNs),
that are logically isolated from each other.
Each tenant has a vSDN controller that controls the
tenant's vSDN.
The hypervisor has the responsibility of ensuring that
each tenant has the impression of controlling the tenant's own vSDN
without interference from the other tenants operating a vSDN on the
same underlying physical SDN network.
The hypervisor is thus essential for amortizing a physical SDN
network installation through offering SDN network services to
multiple users.

We have introduced a main classification of SDN hypervisors according
to their architecture into centralized and distributed hypervisors.
We have further sub-classified the distributed hypervisors according
to their execution platform into hypervisors for
general-purpose computing platforms or for combinations of general-computing
platforms with general- or special-purpose network elements (NEs).
The seminal FlowVisor~\cite{Sherwood2009} has initially been designed with
a centralized architecture and spawned several follow-up designs with a
centralized architecture for both general IP networks as well as
special network types, such as optical and wireless networks.
Concerns about relying on only a single centralized hypervisor, e.g.,
potential overload, have
led to about a dozen distributed SDN hypervisor designs to date.
The distributed hypervisor designs distribute a varying degree of
the hypervisor functions across multiple general-computing platforms
or a mix of general-computing platforms and NEs.
Involving the NEs in the execution of the hypervisor
functions generally led to improved performance and capabilities at the
expense of increased complexity and cost.

There is a wide gamut of important open future research directions
for SDN hypervisors. One important prerequisite for the
future development of SDN hypervisors is a comprehensive performance
evaluation framework.
Informed by our comprehensive review of the existing SDN hypervisors
and their features and limitations,
we have outlined such a performance evaluation framework in
Section~\ref{perf:sec}.
We believe that more research is necessary to refine this framework and
grow it into widely accepted performance benchmarking suite complete
with standard workload traces and test scenarios.
Establishing a unified comprehensive evaluation methodology will likely
provide additional deepened insights into the existing hypervisors
and help guide the research on strategies for advancing the
abstraction and isolation capabilities of the SDN hypervisors while
keeping the overhead introduced by the hypervisor low.

\section*{Acknowledgment}
A first draft of this article was written while Martin Reisslein
visited the Technische Universit\"at M\"unchen (TUM), Germany,
from January through July 2015. 
Support from TUM and a Friedrich Wilhelm Bessel Research Award from
the Alexander von Humboldt Foundation are gratefully acknowledged.

\bibliographystyle{IEEEtran}


\begin{IEEEbiography}{Andreas Blenk}
Andreas Blenk studied computer science at the University of Wu\"rzburg,
Germany, where he received his diploma degree in 2012. After this he
joined the Chair of Communication Networks at the Technische
Universität München in June 2012, where he is working as research and
teaching associate. As a member of the software defined networking and
network virtualization research group, he is currently pursuing his
PhD. His research is focused on service-aware network virtualization,
virtualizing software defined networks, as well as resource management
and embedding algorithms for virtualized networks.
\end{IEEEbiography}

\begin{IEEEbiography} {Arsany Basta}
received his M.Sc. degree in Communication Engineering from the
Technische Universit\"at M\"unchen (TUM) in October 2012. He joined
the TUM Institute of Communication Networks in November 2012 as a
Ph.D. student and a member of the research and teaching staff. His
current research focuses on the applications of Software Defined
Networking (SDN), Network Virtualization (NV), and Network Function 
Virtualization (NFV) to the mobile core towards the next generation (5G) 
network.
\end{IEEEbiography}

\begin{IEEEbiography}{Martin Reisslein} (A'96-S'97-M'98-SM'03-F'14) 
is a Professor in the School of Electrical, Computer, and Energy
Engineering at Arizona State University (ASU), Tempe. He received the
Ph.D. in systems engineering from the University of Pennsylvania in
1998. Martin Reisslein served as Editor-in-Chief for the \textit{IEEE
  Communications Surveys and Tutorials} from 2003--2007 and as
Associate Editor for the \textit{IEEE/ACM Transactions on Networking}
from 2009--2013. He currently serves as Associate Editor for the
\textit{IEEE Transactions on Education} as well as \textit{Computer
  Networks} and \textit{Optical Switching and Networking}.
\end{IEEEbiography}

\begin{IEEEbiography}{Wolfgang Kellerer} (M'96-SM'11) 
is full professor at the Technische Universit\"at M\"unchen (TUM),
heading the Chair of Communication Networks in the Department of
Electrical and Computer Engineering since 2012. Before, he has been
director and head of wireless technology and mobile network research
at NTT DOCOMO's European research laboratories, DOCOMO Euro-Labs, for
more than ten years. His research focuses on concepts for the dynamic
control of networks (Software Defined Networking), network
virtualization and network function virtualization, and 
application-aware traffic management. In the area of wireless networks the
emphasis is on Machine-to-Machine communication, Device-to-Device
communication and wireless sensor networks with a focus on resource
management towards a concept for 5th generation mobile
communications (5G). His research resulted in more than 200
publications and 29 granted patents in the areas of mobile networking
and service platforms. He is a member of the ACM and the VDE ITG, and a Senior
Member of the IEEE.
\end{IEEEbiography}

\end{document}